\documentclass[aps,prd,onecolumn,showpacs,showkeys,superscriptaddress,preprintnumbers,floatfix,reprint]{revtex4}
\usepackage{graphicx}
\usepackage{amsmath}
\usepackage{bm}
\usepackage{yhmath}
\usepackage[colorlinks]{hyperref}
\usepackage{subfigure}
\usepackage{color}
\usepackage{cases}
\usepackage{subfigure}
\usepackage{dcolumn,booktabs,bm}

\usepackage{amsfonts,amssymb,stmaryrd,latexsym}
\usepackage{textcomp}

\usepackage{array} 
\newcounter{RomanNumber}

\newcommand{\lyxmathsym}[1]{\ifmmode\begingroup\def\b@ld{bold}
  \text{\ifx\math@version\b@ld\bfseries\fi#1}\endgroup\else#1\fi}

\def\m{\mathcal}
\def\n{\nonumber}
\def\f{\frac}

\def\p{\partial}

\def\s{\sigma}

\def\a{\alpha}
\def\b{\beta}

\def\L{\Lambda}
\def\l{\lambda}
\def\g{\gamma}
\def\XX{\Xi_{cc}\Xi_{cc}}

\def\X{\Xi_{cc}}

\begin{document}
\title{Deuteron-like states composed of two doubly charmed baryons}

\author{Lu Meng}\email{lmeng@pku.edu.cn}\affiliation{Department of Physics and State Key Laboratory of Nuclear Physics and Technology, Peking University, Beijing 100871, China}

\author{Ning Li}\email{n.li@fz-juelich.de}\affiliation{Institute~for~Advanced~Simulation, Institut~f\"{u}r~Kernphysik,
and
J\"{u}lich~Center~for~Hadron~Physics,~Forschungszentrum~J\"{u}lich,
D-52425~J\"{u}lich, Germany}

\author{Shi-Lin Zhu}\email{zhusl@pku.edu.cn}\affiliation{Department of Physics and State Key Laboratory of Nuclear Physics and Technology, Peking University, Beijing 100871, China}\affiliation{Collaborative Innovation Center of Quantum Matter, Beijing 100871, China}

\begin{abstract}

We present a systematic investigation of the possible molecular
states composed of a pair of doubly charmed baryons
($\Xi_{cc}\Xi_{cc}$) or one doubly charmed baryon and one doubly
charmed antibaryon $(\Xi_{cc}\bar{\Xi}_{cc})$ within the framework
of the one-boson-exchange-potential model. For the spin-triplet
systems, we take into account the mixing between the ${}^3S_1$ and
${}^3D_1$ channels. For the baryon-baryon system $\Xi_{cc}\Xi_{cc}$
with $(R,I) = (\bar{3}, 1/2)$ and $(\bar{3}, 0)$, where $R$ and $I$ represent the group representation and the
isospin of the system, respectively, there exist
loosely bound molecular states. For the baryon-antibaryon system
$\Xi_{cc}\bar{\Xi}_{cc}$ with $(R,I) = (8, 1)$, $(8, 1/2)$ and
$(8,0)$, there also exist deuteron-like molecules. The
$B_{cc}\bar{B}_{cc}$ molecular states may be produced at LHC. The
proximity of their masses to the threshold of two doubly charmed
baryons provides a clean clue to identify them.

\end{abstract}



\maketitle

\thispagestyle{empty}


\section{INTRODUCTION}\label{Sec1}

In 2003, the Belle Collaboration discovered the charmonium-like
state $X(3872)$~\cite{Choi:2003ue}. Subsequently, more
charmonium-/bottomonium-like states such as $Y(4260)$
\cite{Aubert:2005rm}, $Z_c(3900)$
\cite{Ablikim:2013mio,Liu:2013dau}, $Y(4140)$ \cite{Aaltonen:2009tz}
and $Y_b(10888) $\cite{Chen:2008xia} were observed by the BARBAR,
BESIII, Belle, CDF and Belle collaborations respectively. Recently, two
hidden-charm pentaquark states $P_c(4380)$ and $P_c(4450)$ were
observed by the LHCb Collaboration~\cite{Aaij:2015tga}. The
experimental and theoretical progress on the hidden-charm multiquark
states can be found in the recent review \cite{chx}.

It's difficult to accommodate all these XYZ states in the
conventional hadron spectrum. Especially the charged charmonium-like
states are probably good candidates of multiquark states. Some XYZ
states lie very close to the threshold of two charmed hadrons. They
are speculated to be candidates of the hadronic molecular states.

A hadronic molecule is a loosely bound state formed by two
color-singlet hadrons. The molecular states are bound by the
residual strong interaction. For example, the deuteron is a
well-established hadronic molecule, which is a loosely bound state
formed by the proton and neutron. Its binding energy is about 2.225
MeV and root-mean-square radius around 2.0 fm. Compared to the size
of the conventional meson and baryon, the deuteron is really loosely
bound. Besides the deuteron, Voloshin and Okun investigated the
possible molecular states formed by a charmed meson and a charmed
antimeson forty years ago \cite{Voloshin:1976ap}. Also, De Rujula
{\it et al} tried to explain $\psi(4040)$ as a $D^*\bar{D^*}$
molecular state in \cite{DeRujula:1976zlg}. In
\cite{Tornqvist:1993ng,Tornqvist:1991ks}, T\"ornqvist analysed the
possible deuteron-like $D\bar{D}^*$ and $D^*\bar{D}^*$ molecules.
In literature, there are many investigations on the hadronic molecules
such as the $\Lambda(1405)$ as a candidate of the $\bar{K}N$ molecule \cite{Mai:2012dt,Hall:2014uca},
the dibaryon composed of two light baryons
~\cite{Straub:1990de,Huang:2004ke,Dai:2006gs,Zhang:2006dy,Chen:2007qn,Ping:2008tp,Huang:2011kf,Chen:2011zzb}, the possible
molecular states composed of a pair of heavy
mesons~\cite{Liu:2007bf,Liu:2008xz,Liu:2008tn,Liu:2009ei,Liu:2008fh,Zhao:2015mga,Zhao:2014gqa},
the molecular states composed of a pair of heavy
baryons~\cite{Lee:2011rka,Li:2012bt,Vijande:2016nzk,Carames:2015sya,Huang:2013rla,Gerasyuta:2011zx},
$\Lambda_c \Lambda_c$ and $\Lambda_c N$ bound
states~\cite{Liu:2011xc,Meguro:2011nr} and the possible bound states
of $\Sigma_c N$, $\Xi_c'N$, $\Xi_{cc}N$, $\Xi\Xi_{cc}$,
~\cite{Froemel:2004ea,JuliaDiaz:2004rf}.

Very recently, many events with four heavy quarks
($QQ\bar{Q}\bar{Q}$) were reported by different collaborations. For
example, the $J/\psi$ pairs were observed by LHCb \cite{Aaij:2011yc}
and CMS collaborations\cite{Khachatryan:2014iia}. The simultaneous
$J/\psi\Upsilon(1S)$ events were reported by both
D0\cite{Abazov:2015fbl} and CMS\cite{Kamuran}. CMS Collaboration
also observed the simultaneous $\Upsilon(1S)\Upsilon(1S)$
events\cite{Maksat}. Some of these $QQ\bar{Q}\bar{Q}$ events may be
resonant. There are extensive theoretical discussions about the
possible $QQ\bar{Q}\bar{Q}$ states
\cite{Brambilla:2015rqa,Chen:2016jxd,Wu:2016vtq,Bai:2016int,Karliner:2016zzc,Wang:2017jtz}.

In this work, we investigate the possible deuteron-like hadronic
molecules composed of two doubly charmed baryons.
These states have the configurations such as $\Xi_{cc}\Xi_{cc}$ or
$\Xi_{cc}\bar{\Xi}_{cc}$. Especially, the possible
$\Xi_{cc}\bar{\Xi}_{cc}$ molecular states can be searched for at
LHC. We will adopt the one-boson-exchange-potential model (OBEP).
Aside from the long-range $\pi$ exchange force
~\cite{Yukawa:1935xg}, the OBEP model also introduces the
medium-range $\sigma$ exchange as well as the short-range $\rho$ and
$\omega$ exchange forces.

We organize the paper as follows. After the introduction, we present
the theoretical formalism including the Lagrangians, the derivations
of the coupling constants and the interaction potential in
Section~\ref{Secform}. Our numerical results are given in
Section~\ref{Secno}. We summarize our results and make some
discussions in Section~\ref{Seccnclsn}. Some useful formulae are
collected in the Appendix.

\section{FORMALISM}\label{Secform}

In 2002, the SELEX Collaboration reported a doubly charmed state
with mass $3520$ MeV~\cite{Mattson:2002vu}. This structure contains
two charm quarks and a down quark, and it is denoted by $\Xi_{cc}$.
Later this state was confirmed by the same
collaboration~\cite{Ocherashvili:2004hi}. In a conference
report~\cite{Engelfried:2007at}, another state containing two charm
quarks and an up quark at $3780$ MeV was reported also by the SELEX
Collaboration. In Refs.
\cite{Martynenko:2007je,Shah:2017liu,Karliner:2014gca,Yoshida:2015tia,Sun:2014aya,Sun:2016wzh},
the mass of the doubly charmed baryon was estimated from $3511$ to
$3685$ MeV. The particular isospin splitting of the states observed by
   SELEX was discussed in Ref.~\cite{Brodsky:2011zs}.

The doubly charmed baryon $\Xi_{cc}$ is composed of two charm quarks
and one light quark. The wave function of the two charm quarks is
\begin{eqnarray}
\psi_{cc} = \psi^{flavor}_{cc} \otimes \psi^{color}_{cc} \otimes
\psi^{spin}_{cc}\otimes \psi^{space}_{cc}.
\end{eqnarray}
For the ground state, both the flavor wave function
$\psi^{flavor}_{cc}$ and space wave function $ \psi^{space}_{cc}$
are symmetric while its color wave function $\psi^{color}_{cc}$ is
antisymmetric under the exchange of the two charm quarks. Hence, the
spin wave function $\psi_{cc}^{spin}$ is symmetric as required by
Pauli Principle, ie., $S_{cc} = 1$. As a result, the spin of
$\Xi_{cc}$ for the ground state is $\frac{1}{2}$ or $\frac{3}{2}$.
In the present work, we focus on the molecular systems composed of
two spin-$\frac{1}{2}$ $\Xi_{cc}$. They should be the lightest
states among molecular states with various spin configurations.

The heavy charm quarks act as the static color source. The doubly
charmed baryons form the fundamental representation in the SU(3)
flavor space regarding to the light quarks. For convenience, we
adopt the notation, $B_{cc} = \left(\Xi_{cc}^{u}, \Xi_{cc}^{d},
\Xi_{cc}^{s} \right)^T$, where the superscripts of $\Xi_{cc}$ denote the
corresponding light quarks and superscript $T$ means transpose of the matrix.

Under the SU(3)-flavor symmetry, the $B_{cc}B_{cc}$ systems are
decomposed as $3_{F} \otimes 3_{F} = 6_{F} \oplus \bar{3}_{F}$ while
the $B_{cc}\bar{B}_{cc}$ systems can be decomposed as $3_{F} \otimes
\bar{3}_{F} = 8_F \oplus 1_F$. For simplicity, we use $(R, I)$ to
denote the systems, where $R$ and $I$ represent the group
representation and isospin, respectively. The relevant flavor wave
functions are given in Table~\ref{Table:flavor}.

\begin{table}[htp]
 \centering
 \caption{Flavor wave functions of the $B_{cc}B_{cc}$ and
$B_{cc}\bar{B}_{cc}$ systems. $R$ and $I$ denotes the group
representation and the isospin respectively.}\label{Table:flavor}
\begin{tabular}{cc|cc|cc}
\toprule[1pt]\toprule[1pt] Systems/$(R,I)$ & Flavor &
Systems/$(R,I)$  & Flavor & Systems/$(R,I)$  & Flavor
\tabularnewline \midrule[1pt] $(6,1)$                 & $uu$ &
$(\bar{3},\frac{1}{2})$ & $\frac{1}{\sqrt{2}}\left(us-su\right)$ &
$(8,\frac{1}{2})$       & $u\bar{s}$\tabularnewline
                        & $\frac{1}{\sqrt{2}}\left(ud+du\right)$ &
                        & $\frac{1}{\sqrt{2}}\left(ds-sd\right)$ &
                        & $d\bar{s}$\tabularnewline
                        & $dd$ &
$(\bar{3},0)$           & $\frac{1}{\sqrt{2}}\left(ud-du\right)$ &
$(8,\frac{1}{2})$       & $s\bar{d}$\tabularnewline
$(6,\frac{1}{2})$       & $\frac{1}{\sqrt{2}}\left(us+su\right)$&
$(8,1)$                 & $u\bar{d}$ &
                        & $s\bar{u}$\tabularnewline
                        & $\frac{1}{\sqrt{2}}\left(ds+sd\right)$ &
                        & $\frac{1}{\sqrt{2}}\left(u\bar{u}-d\bar{d}\right)$ &
$(8,0)$                 &
$\frac{1}{\sqrt{6}}\left(u\bar{u}+d\bar{d}-2s\bar{s}\right)$\tabularnewline
$(6,0)$                 & $ss$ &
                        & $d\bar{u}$ &
$(1,0)$                 &
$\frac{1}{\sqrt{3}}\left(u\bar{u}+d\bar{d}+s\bar{s}\right)$\tabularnewline
\bottomrule[1pt]\bottomrule[1pt]
\end{tabular}
\end{table}

\subsection{The Lagrangian}\label{sec:lagrangian}

The notations for the exchanged pseudoscalar and vector mesons read
\begin{eqnarray}
\mathcal{M} = \left(
\begin{array}{ccc}
\frac{\pi^0}{\sqrt{2}}+\frac{\eta}{\sqrt{6}}&     \pi^+               &  K^+  \\
\pi^-                   &-\frac{\pi^0}{\sqrt{2}}+\frac{\eta}{\sqrt{6}}&  K^0   \\
K^-                  &\bar{K}^0                    &-\frac{2}{\sqrt{6}}\eta  \\
\end{array}
\right),\quad \mathcal{V}^{\mu} = \left(
\begin{array}{ccc}
\frac{\rho^0}{\sqrt{2}}+\frac{\omega}{\sqrt{2}}&     \rho^{+}               &  K^{*+}  \\
\rho^-                    &-\frac{\rho^0}{\sqrt{2}}+\frac{\omega}{\sqrt{2}} &  K^{*0}   \\
K^{*-}                  &\bar{K}^{*0}                    & \phi  \\
\end{array}
\right)^{\mu}. \label{light:mesons}
\end{eqnarray}
Some heavier-meson exchanges which provide very short-range interactions are not included
since we focus on the very loosely bound states.
Under the SU(3)-flavor symmetry, we construct the Lagrangian for the
pseudoscalar exchange as
\begin{equation}
\m{L}_{phh}=g_{phh}\bar{B}_{cc}i\g_5 \mathcal{M} B_{cc}.
\label{lagrangian:p}
\end{equation}
One may also use the axial-vector coupling,
\begin{eqnarray}
\m{L}_{phh}=f_{phh}\bar{B}_{cc}\gamma_5\gamma_{\mu} \partial^{\mu}
\mathcal{M} B_{cc}, \label{lagrangian:pv}
\end{eqnarray}
The above two Lagrangians are equivalent at the tree level. In the
current calculation, we adopt Eq.~(\ref{lagrangian:p}). For the
vector-meson exchange, we have
\begin{equation}
\m{L}_{vhh}=g_{vhh}\bar{B}\g_{\mu}\mathcal{V}^{\mu}B_{cc}
+\f{f_{vhh}}{2m}\bar{B}\s_{\mu\nu}\p^{\mu}\mathcal{V}^{\nu}B_{cc}, \label{lagrangian:v}\\
\end{equation}
and for the scalar-meson exchange,
\begin{equation}
\m{L}_{\s hh}=g_{\s hh}\bar{B}_{cc}\s B_{cc}.   \label{lagrangian:s}
\end{equation}
In the previous expressions, $g_{phh}$, $g_{vhh}$, $f_{vhh}$ and
$g_{\sigma hh}$ are the coupling constants. Their values are given
in Section~\ref{sec:coupling}.

\begin{table}
 \centering
 \caption{The Coupling constants and the masses of the relevant
hadrons~\cite{Machleidt:2000ge,Cao:2010km,Olive:2016xmw,Machleidt:1987hj}.
For the pion and kaon multiplets, their averaged masses are used.
$m_{\X}$ is the mass of $\Xi_{cc}^+$ reported in Ref.
\cite{Ocherashvili:2004hi,Mattson:2002vu}.}
\label{Table:masscoupling}
\begin{tabular}{cc|cc|cc|cc|cc}
\toprule[1pt]\toprule[1pt] Baryons & Mass (MeV) & Mesons & Mass
(MeV) & Mesons & Mass (MeV) & Coupling  & Value & Coupling &
Value\tabularnewline \midrule[1pt] $\Xi_{cc}^{u,d,s}$ & 3520 & $\pi$
& 137.27 & $\phi$ & 1019.46 & $g_{\pi NN}^{2}/4\pi$ & 13.6 &
$g_{phh}$ & -13.86\tabularnewline Proton ($p$) & 938.27 & $\eta$ &
547.85 & $K$ & 495.65 & $g_{\rho NN}^{2}/4\pi$ & 0.84 & $g_{vhh}$ &
4.60\tabularnewline Neutron ($n$) & 939.57 & $\rho$ & 775.49 &
$K^{*}$ & 893.80 & $f_{\rho NN}/g_{\rho NN}$ & 6.1 & $f_{vhh}$ &
-29.06\tabularnewline
 &  & $\omega$ & 782.65 & $\sigma$ & 600 & $g_{\sigma NN}^{2}/4\pi$ & 5.69 & $g_{\sigma hh}$ & 2.82 \tabularnewline
\bottomrule[1pt]\bottomrule[1pt]
\end{tabular}
\end{table}

\subsection{Coupling Constants}\label{sec:coupling}

In this subsection, we focus on the derivation of the coupling
constants used in the current work. The coupling constants for the
light bosons interacting with the nucleon are relatively well-known.
They can either be extracted from experimental data or calculated
from various models. We will derive the values of the coupling
constants with the help of the quark model. We denote the coupling
constants between the light mesons and the doubly charmed baryons as
$g_{mB_{cc}B_{cc}}$, those between the light mesons and the quarks
as $g_{mqq}$, and those between the light mesons and the nucleon as
$g_{mNN}$. We make use of the relations as
follows,
\begin{eqnarray}
\langle  p\uparrow |\mathcal{L}_{mNN}|p\uparrow \rangle &=& \langle p\uparrow |\mathcal{L}_{mqq}|p\uparrow \rangle, \label{cp:nucl}  \\
\langle \Xi_{cc}^u\uparrow |\mathcal{L}_{mhh}|\Xi_{cc}^u
\uparrow\rangle &=& \langle \Xi_{cc}^u\uparrow
|\mathcal{L}_{mqq}|\Xi_{cc}^u\uparrow \rangle.\label{cp:Bcc}
\end{eqnarray}
where ``$\uparrow$" means the third component of the spin is $+1/2$.
The matrix elements are calculated both at hadron and quark level
respectively. We first derive the relation between $g_{mqq}$ and
$g_{mNN}$ from Eq. (\ref{cp:nucl}), and then obtain the relation
between $g_{mB_{cc}B_{cc}}$ and $g_{mqq}$ from Eq. (\ref{cp:Bcc}).
Both relations contain quark masses. Finally, we combine the two
relations and obtain the relation between $g_{mB_{cc}B_{cc}}$ and
$g_{mNN}$ without the quark mass dependence.

At the hadron level, the Lagrangians for the light mesons and the
nucleon are
\begin{eqnarray}
\mathcal{L}_{\pi NN}&=&g_{\pi
NN}\bar{N}i\gamma_{5}\bm{\tau}\cdot\bm{\pi}N , \label{lagrangian:npi} \\
 \mathcal{L}_{\rho NN}&=&g_{\rho
NN}\bar{N}\gamma_{\mu}\bm{\tau}\cdot\bm{\rho}^{\mu}N +\frac{f_{\rho
NN}}{2m_{N}}\bar{N}\sigma_{\mu\nu}\bm{(\tau}\cdot\partial^{\mu}\bm{\rho}^{\nu})N, \label{lagrangian:nrho}  \\
\mathcal{L}_{\sigma NN}&=&g_{\sigma
NN}\bar{N}\sigma N, \label{lagrangian:nsigma}
\end{eqnarray}
where $N = (p, n)^T$ with $p$ and $n$ the proton and neutron
respectively. The numerical values of the coupling constants,
$g_{\pi NN}$, $g_{\rho NN}$, $f_{\rho NN}$ and $g_{\sigma NN}$ are
taken from Refs.
~\cite{Cao:2010km,Machleidt:2000ge,Machleidt:1987hj} and collected
in Table~\ref{Table:masscoupling}.

At the quark level, the Lagrangian reads
\begin{equation}
\mathcal{L}_{q}=g_{pqq}\bar{q}i\gamma_{5}\mathcal{M}q+g_{vqq}\bar{q}\gamma_{\mu}\mathcal{V}^{\mu}q+g_{\sigma
qq}\bar{q}\sigma q
\end{equation}
where $q=(u,d,s)^T$ is the light quark triplet. Notice that in the
above expression we do not consider the tensor part as we do at the
hadron level (the second part of the Eq.~(\ref{lagrangian:nrho}))
for the vector-meson exchange because the quarks are taken as point
particles whereas the hadrons are not.

The amplitudes for the two baryons
and $\pi^0$ vertices read,
\begin{eqnarray}
i\mathcal{M}_{\pi^{0}p\uparrow p\uparrow}&=&g_{\pi NN}\frac{Q_{3}}{m_{N}}=\frac{1}{\sqrt{2}}g_{pqq}\frac{Q_{3}}{m_{q}}\times\frac{5}{3},\\
i\mathcal{M}_{\pi^{0}\Xi_{cc}^{u}\uparrow\Xi_{cc}^{u}\uparrow}&=&\frac{1}{\sqrt{2}}g_{phh}\frac{Q_{3}}{m_{\Xi_{cc}}}=\frac{1}{\sqrt{2}}g_{pqq}\frac{Q_{3}}{m_{q}}\times\left(-\frac{1}{3}\right),
\end{eqnarray}
where $m_{q}$, $m_{N}$ and $m_{\Xi_{cc}}$ are the masses of the
quark, nucleon and doubly charmed baryon respectively while $Q_3$ is
the third component of the pion momentum. With the above relation,
one obtain $g_{phh}$ directly.
Finally, we obtain the
all coupling constants used in the current work as
\begin{align}
g_{\sigma hh}&=\frac{1}{3}g_{\sigma NN}, & g_{phh}&=-\frac{\sqrt{2}}{5}\frac{m_{\X}}{m_{N}}g_{\pi NN},\\
g_{vhh}&=\sqrt{2}g_{\rho NN}, &
g_{vhh}+f_{vhh}&=-\frac{\sqrt{2}}{5}\left(g_{\rho NN}+f_{\rho
NN}\right)\frac{m_{\X}}{m_{N}},
\end{align}
 For the vector-meson exchange, we use the
values of $g_{\rho NN}$ but not $g_{\omega NN}$ because $g_{\rho
NN}$ is more stable than $g_{\omega NN}$ in different models. The
numerical values of the coupling constants are given in
Table~\ref{Table:masscoupling}. For the doubly charmed baryon
masses, we assume the exact SU(3)-flavor symmetry and take the
results from the SELEX Collaboration~\cite{Mattson:2002vu}, 3520
MeV, for all the doubly charmed baryons covered in the work.

\subsection{The Interaction Potentials}

With the Lagrangians in Section~\ref{sec:lagrangian}, we derive the
interaction potentials in momentum space. Due to the large masses of the doubly charmed baryons, 
the interaction potential in the momentum space $V(\bm Q)$ is expanded in terms of $\bm{Q}/m_{\Xi_{cc}}$,  or $\bm{k}/m_{\Xi_{cc}}$, 
where $\bm{Q}$ is $(\bm{p}_f-\bm{p}_i)$ while $\bm{k}$ is $(\bm{p}_i+\bm{p}_j)/2$, and kept up to order 
$\mathcal{O}(\bm{Q}^2/m^2_{\Xi_{cc}}, \bm{k}^2/m_{\Xi_{cc}}^2)$. In our case, $Q_0^2$ is in fact a  high order term and 
can be neglected directly, see Appendix \ref{app_func} for a short analysis of $Q_0^2$. After transforming the potential into the 
coordinate space, the conjugate variable of $\bm{Q}$ is $\bm{r}$ and that of $\bm{k}$ is $-i\nabla$. The latter provides the 
only nonlocal potential in the present calculations, i.e. the spin-orbit force. Other nonlocal interactions such as the recoil effect are 
neglected. It is mentioned in Ref. \cite{Machleidt:2000ge} 
that the nonlocal potential changes the off-shell behavior. However, in the present work we are mainly interested in the hadronic 
molecular states composed of the doubly charmed baryons, in which the bounded hadrons are approximately on-shell. Hence, 
it is reasonable to neglect the nonlocal potential other than the spin-orbit force in our calculation.

When performing the Fourier
Transformation, we introduce a monopole form factor,
\begin{eqnarray}
\m{F}(\bm Q)=\f{\L^2-m_{ex}^2}{\L^2-Q^2}=\f{\L^2-m_{ex}^2}{\l^2+\bm{Q}^2},\label{FF}
\end{eqnarray}
for each vertex. $\Lambda$ is a cutoff parameter, which is used to
suppress the high-momenta contribution or equivalently, to soften
the short-range interactions. $m_{ex}$ and $Q$ are the mass and four
momentum of the exchanged meson respectively, and
$\l^2={\L}^2-Q_0^2$. After the Fourier Transformation,
\begin{eqnarray}
\m{V}(r)=\f{1}{(2\pi)^3}\int d\bm{Q}
e^{i\bm{Q}\cdot\bm{r}}\m{V}(\bm Q)\mathcal{F}^2(\bm Q),
\end{eqnarray}
one obtains the interaction potentials in coordinate space which
read
\begin{itemize}

\item Pseudoscalar exchange:
\begin{eqnarray}
\m{V}_{SS}^p(r;\a)&=&C_{\a}^p\f{g_{1p}g_{2p}}{4\pi}\f{m_{\a}^3}{12m_{\Xi_{cc}}^2}
H_1(\L,m_{\a},r)\bm{\s}_1\cdot\bm{\s}_2, \n\\
\m{V}_T^p(r;\a)&=&C_{\a}^p\f{g_{1p}g_{2p}}{4\pi}\f{m_{\a}^3}{12m_{\Xi_{cc}}^2}
H_3(\L,m_{\a},r)S_{12}(\hat{r}), \label{potential:p1}
\end{eqnarray}

\item Vector exchange:
\begin{eqnarray}
\m{V}_C^v(r;\b)&=&C_{\b}^v\f{m_{\b}}{4\pi}\left[
g_{1v}g_{2v}H_0(\L,m_{\b},r)
+\f{m_{\b}^2}{8m_{\Xi_{cc}}^2}(g_{1v}g_{2v}+2g_{1v}f_{2v}+2g_{2v}f_{1v})H_1(\L,m_{\b},r)\right], \n\\
\m{V}_{SS}^v(r;\b)&=&C_{\b}^v\left[g_{1v}g_{2v}+g_{1v}f_{2v}+g_{2v}f_{1v}+f_{1v}f_{2v}\right]
\f{1}{4\pi}\f{m_{\b}^3}{6m_{\Xi_{cc}}^2}H_1(\L,m_{\b},r)\bm{\s}_1\cdot\bm{\s}_1, \n\\
\m{V}_T^v(r;\b)&=&-C_{\b}^v\left[g_{1v}g_{2v}+g_{1v}f_{2v}+g_{2v}f_{1v}+f_{1v}f_{2v}\right]
\f{1}{4\pi}\f{m_{\b}^3}{12m_{\Xi_{cc}}^2}H_3(\L,m_{\b},r)S_{12}(\hat{r}), \n\\
\m{V}_{LS}^v(r;\b)&=&-C_{\b}^v\f{1}{4\pi}\f{m_{\b}^3}{2m_{\Xi_{cc}}^2}H_2(\L,m_{\b},r)
\left[3g_{1v}g_{2v}\bm{L}
\cdot\bm{S}+4g_{2v}f_{1v}\bm{L}\cdot\bm{S}_1
+4g_{1v}f_{2v}\bm{L}\cdot\bm{S}_2\right], \label{potential:v}
\end{eqnarray}
\item Scalar exchange:
\begin{eqnarray}
\m{V}_C^s(r;\s)&=&-C_{\s}^sm_{\s}\f{g_{1s}g_{2s}}{4\pi}\left[H_0(\L,m_{\s},r)
-\f{m_{\s}^2}{8m_{\Xi_{cc}}^2}H_1(\L,m_{\s},r)\right],\n\\
\m{V}_{LS}^s(r;\s)&=&-C_{\s}^s\f{g_{1s}g_{2s}}{4\pi}\f{m_{\s}^3}{2m_{\Xi_cc}^2}
H_2(\L,m_{\s},r)\bm{L}\cdot\bm{S}.\label{potential:s}
\end{eqnarray}
\end{itemize}
In the above expressions, the superscripts $p$, $s$ and $v$ denote
the pseudoscalar, scalar and vector mesons, respectively. $\alpha =
\pi$, $\eta$ or $K$ while $\beta = \omega$, $\rho$, $\phi$ and
$K^*$. The specific expressions of the scalar functions $H_0$,
$H_1$, $H_2$ and $H_3$ are given in Appendix \ref{app_func}. Some
details about the so-called "contact interaction" are also included
in Appendix \ref{app_func}. $C_\alpha^p$, $C_\beta^v$ and
$C_\sigma^s$ are the isospin factors. Their numerical values are
given in Table~\ref{Table:Isospinfac}. $\bm{L}$ is the relative
orbit angular momentum operator between the two baryons while
$\bm{S}_{1(2)}$ is the spin operator for baryon $1(2)$. The total
spin operator of the two-baryon system is $\bm{S} = \bm{S}_1 +
\bm{S}_2$.  $S_{12}(\hat{r})=3(\bm{\sigma}_1 \cdot
\hat{r})(\bm{\sigma}_2 \cdot \hat{r}) - \bm{\sigma}_1 \cdot
\bm{\sigma}_2 $ is the tensor operator which mixes the $S$- and
$D$-waves.

With the specific expressions in Eqs.
~(\ref{potential:p1}-\ref{potential:s}) and  the isospin factors
given in Table~\ref{Table:Isospinfac}, one can obtain the potentials
for the $B_{cc}B_{cc}$ systems. Instead of calculating Feynman amplitude of tree diagram, we can use the "G-parity" rule to
derive the potentials of the $B_{cc}\bar{B}_{cc}$ systems directly
from the potentials for the $B_{cc}B_{cc}$ systems if the exchanged meson has certain "G-parity". For example, one
immediately obtains the pion-exchange potential for the
$B_{cc}\bar{B}_{cc}$ system with $(R,I)= (8,1)$ by multiplying the
corresponding potential for the $B_{cc}B_{cc}$ system with $(R,I) =
(6, 1)$ by an factor $(-1)^{G_{\pi}}$ where $G_\pi$ is the
"G-parity" of the pion, see Table~\ref{Table:Isospinfac}.
For the baryon-antibaryon systems, some annihilation potentials
corresponding to the very short-range interactions are not included in the current
calculation since we focus on the study of the loosely bound states.

\begin{table}
  \centering
 \caption{The isospin factors. $R$ and $I$ denote the group
representation and isospin respectively. The left panel is for the
$B_{cc}B_{cc}$ system while the right panel is for the
$B_{cc}\bar{B}_{cc}$ system. }\label{Table:Isospinfac}
\begin{tabular}{ccccccccc|ccccccccc}
\toprule[1pt]\toprule[1pt] Systems/$(R,I)$  & $C_{\pi}^{p}$ &
$C_{\eta}^{p}$ & $C_{K}^{p}$ & $C_{\rho}^{v}$ & $C_{\omega}^{v}$ &
$C_{\phi}^{v}$ & $C_{K^{*}}^{v}$ &
$C_{\text{\ensuremath{\sigma}}}^{s}$& Systems/$(R,I)$ &
$C_{\pi}^{p}$ & $C_{\eta}^{p}$ & $C_{K}^{p}$ & $C_{\rho}^{v}$ &
$C_{\omega}^{v}$ & $C_{\phi}^{v}$ & $C_{K^{*}}^{v}$ &
$C_{\text{\ensuremath{\sigma}}}^{s}$ \tabularnewline \midrule[1pt]
$(6,1)$ & $\frac{1}{2}$ & $\frac{1}{6}$ & $0$ & $\frac{1}{2}$ &
$\frac{1}{2}$ & $0$ & $0$ & $1$& $(8,1)$ & $-\frac{1}{2}$ &
$\frac{1}{6}$ & $0$ & $\frac{1}{2}$ & $-\frac{1}{2}$ & $0$ & $0$ &
$1$\tabularnewline $(6,\frac{1}{2})$ & $0$ & $-\frac{1}{3}$ & $1$ &
$0$ & $0$ & $0$ & $1$ & $1$& $(8,\frac{1}{2})$ & $0$ &
$-\frac{1}{3}$ & $0$ & $0$ & $0$ & $0$ & $0$ & $1$\tabularnewline
$(6,0)$ &$0$& $\frac{2}{3}$ & $0$ & $0$ & $0$ & $1$ & $0$ & $1$&
$(8,\frac{1}{2})$ & $0$ & $-\frac{1}{3}$ & $0$ & $0$ & $0$ & $0$ &
$0$ & $1$\tabularnewline $(\bar{3},\frac{1}{2})$  & $0$ &
$-\frac{1}{3}$ & $-1$ & $0$ & $0$ & $0$ & $-1$ & $1$& $(8,0)$ &
$\frac{1}{2}$ & $\frac{1}{2}$ & $-\frac{4}{3}$ & $-\frac{1}{2}$ &
$-\frac{1}{6}$ & $-\frac{2}{3}$ & $\frac{4}{3}$ & $1$\tabularnewline
$(\bar{3},0)$ & $-\frac{3}{2}$ & $\frac{1}{6}$ & $0$ &
$-\frac{3}{2}$ & $\frac{1}{2}$ & $0$ & $0$ & $1$& $(1,0)$ & $1$ &
$\frac{1}{3}$ & $\frac{4}{3}$ & $-1$ & $-\frac{1}{3}$ &
$-\frac{1}{3}$ & $-\frac{4}{3}$ & $1$\tabularnewline
\bottomrule[1pt]\bottomrule[1pt]
\end{tabular}
\end{table}

Since we focus on the system composed of a pair of
spin-$\frac{1}{2}$ particles, the total spin of the system can be
$0$ or $1$. For the spin-$0$ case, we focus on the ${}^1S_0$ channel
while for the spin-$1$ case we must deal with the ${}^3S_1$ and
${}^3D_1$ simultaneously because of the tensor potential. The wave
functions of the spin-singlet channel read
\begin{eqnarray}
\Psi(r,\theta,\phi)\chi_{ss_z}=y_S(r)|^1S_0\rangle, \label{wave:function:1S0}
\end{eqnarray}
while the wave functions of the spin-triplet channels are
\begin{eqnarray}
\Psi(r,\theta,\phi)^T\chi_{ss_z}^T= \left(\begin{array}{c}
T_S(r) \\
 0     \\
\end{array}\right) |^3S_1\rangle +
\left(\begin{array}{c}
0  \\
T_D(r)\\
\end{array}\right)|^3D_1\rangle,\label{wave:function:3SD}
\end{eqnarray}
In Eq.~(\ref{wave:function:1S0}), $y_S(r)$ is the radial wave
function for the ${}^1S_0$ channel while $T_S^T(r)$ and $T_D^T$ in
Eq.~(\ref{wave:function:3SD}) are the radial wave functions for
${}^3S_1$ and ${}^3D_1$ channels, respectively. For the matrices of
the operators appearing in
Eqs.~(\ref{potential:p1}-\ref{potential:s}), we have
\begin{itemize}
\item Spin-singlet ($S=0$):
\begin{eqnarray}
\bm{\s}_1\cdot\bm{\s}_2=-3,\quad
\bm{L}\cdot\bm{S}=0,\quad\bm{L}\cdot\bm{S}_1=0,
\quad\bm{L}\cdot\bm{S}_2=0,\quad S_{12}(\hat{r})=0,
\label{operator:singlet}
\end{eqnarray}

\item Spin-triplet ($S=1$):
\begin{eqnarray}
\bm{\s}_1\cdot\bm{\s}_2 &=& \left(\begin{array}{cc}
 1  &   0  \\
 0  &   1  \\
 \end{array} \right),~~
 S_{12}(\hat{r})=
 \left(\begin{array}{ccc}
0   &  \sqrt{8}  \\
\sqrt{8}  &  -2   \\
\end{array}\right),~~
\bm{L}\cdot\bm{S}= \left(\begin{array}{cc}
0  &  0 \\
0  & -3 \\
\end{array}\right),~~ \\
 \bm{L}\cdot \bm{S}_1 &=&
 \left(\begin{array}{cc}
 0  &   0  \\
 0  &  -\f{3}{2} \\
 \end{array}\right),~~
 \bm{L}\cdot \bm{S}_2=
 \left( \begin{array}{ccc}
 0   &    0 \\
 0   &  -\f{3}{2} \\
\end{array}\right). \label{operator:triplet}
\end{eqnarray}
\end{itemize}
One may find the details in deriving these matrices in
Appendix~\ref{app_matrix}.

\section{Numerical results }\label{Secno}

We solve the Schr{\"o}dinger equation with the potential derived
before and obtain the binding energy (B. E. ) and the radial wave
function. With the wave functions we also calculate the
root-mean-square radius $r_{rms}$. The root-mean-square radius reads
\begin{eqnarray}
r_{rms}^2 &=& \int y_S^*(r)y_S(r) r^4 dr ,
\end{eqnarray}
for the spin-singlet channels and
\begin{eqnarray}
r_{rms}^2 &=&  \int \left[ T_S^*(r) T_S(r) + T_D^*(r)
T_D(r)\right]r^4 dr,
\end{eqnarray}
for the spin-triplet channels. For the coupled channels, we also
calculate the individual probability for each channel,
\begin{eqnarray}
P_{^3S_1} = \int  T_S^*(r) T_S(r)r^2 dr,
\end{eqnarray}
for the ${}^3S_1$ channel and
\begin{eqnarray}
P_{^3D_1} = \int  T_D^*(r) T_D(r)r^2 dr,
\end{eqnarray}
for the ${}^3D_1$ channel.

In our calculation, we need the value of the cutoff. The study of
the deuteron with the OBEP model suggests a reasonable range for the
cutoff, $0.80 - 1.50$ GeV. Since the doubly charmed baryon is much
heavier than the nucleon, we take a slightly wider range $0.8 - 2.0$
GeV for the cutoff parameter.

\subsection{$B_{cc}B_{cc}$ systems }

For the $B_{cc}B_{cc}$ systems, the total wave functions should be
antisymmetric under exchange of the two baryons, required by Pauli
Principle. Given that the spacial wave functions are symmetric ($S$
or $D$ waves), the spin of the system is $1$ and 0 for the
$\bar{3}$-representation and $6$-representation respectively.

\subsubsection{$\bar{3}$-representation, $S=1$}

\begin{figure}[htp]
\centering
\begin{tabular}{ccc}
\includegraphics[width=0.30\textwidth]{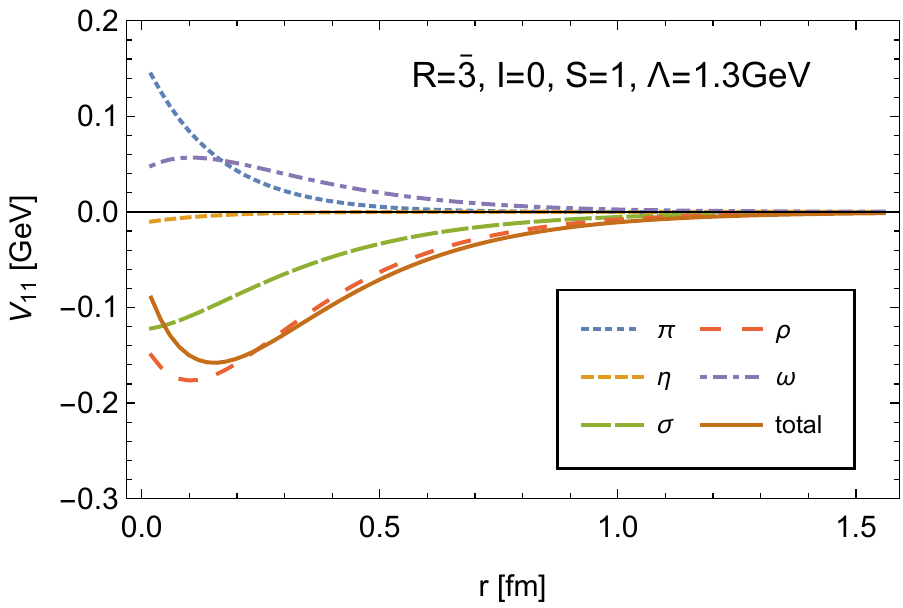}~&~
\includegraphics[width=0.30\textwidth]{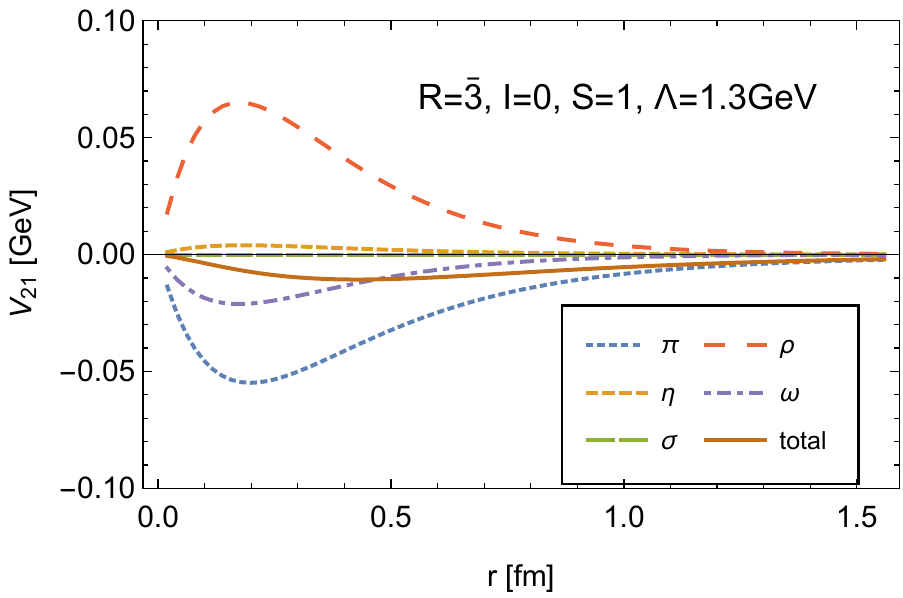}~&~
\includegraphics[width=0.30\textwidth]{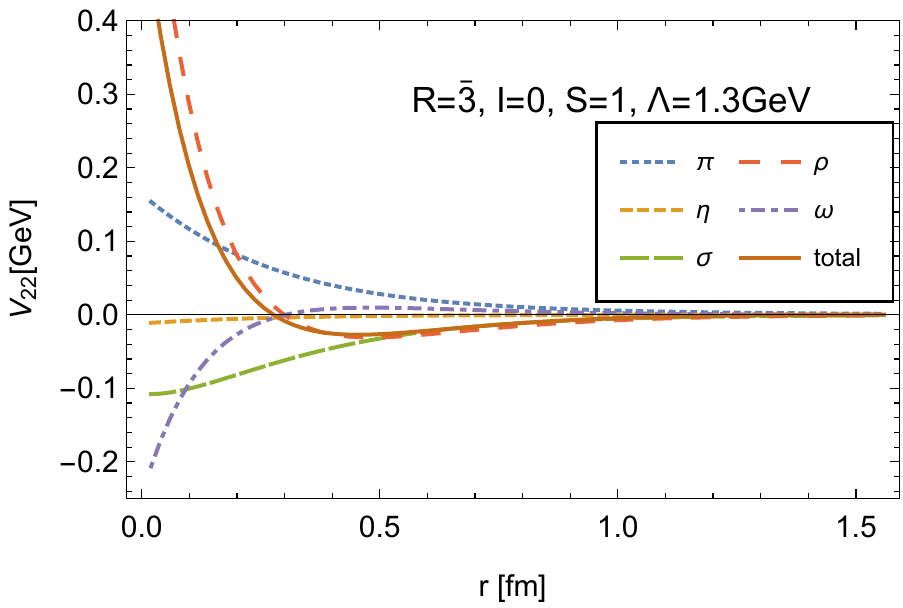}\\
\includegraphics[width=0.30\textwidth]{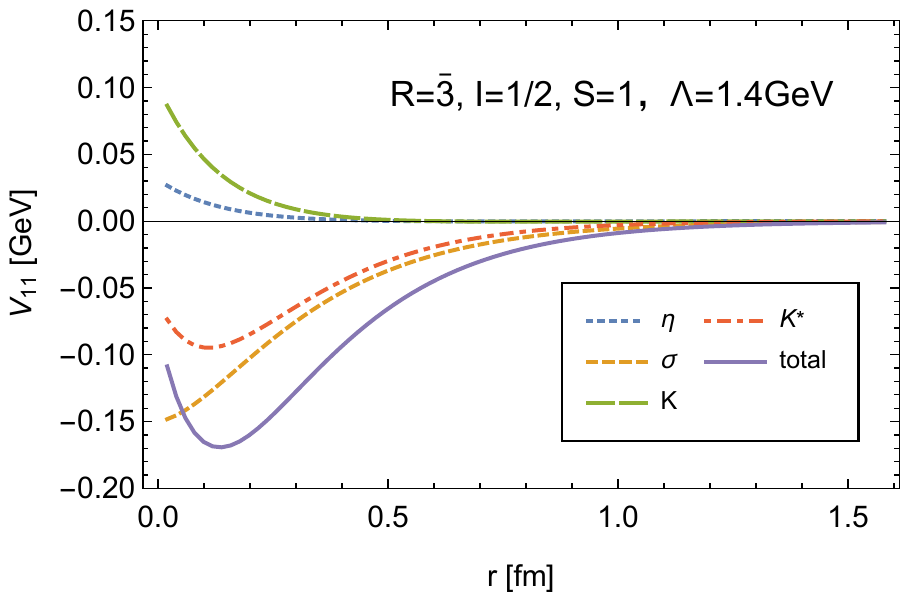}~&~
\includegraphics[width=0.30\textwidth]{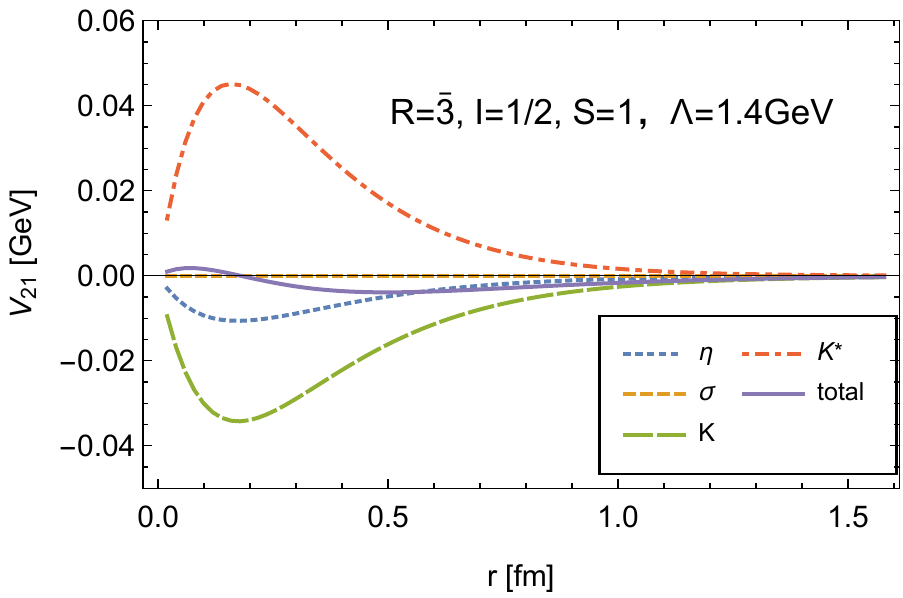}~&~
\includegraphics[width=0.30\textwidth]{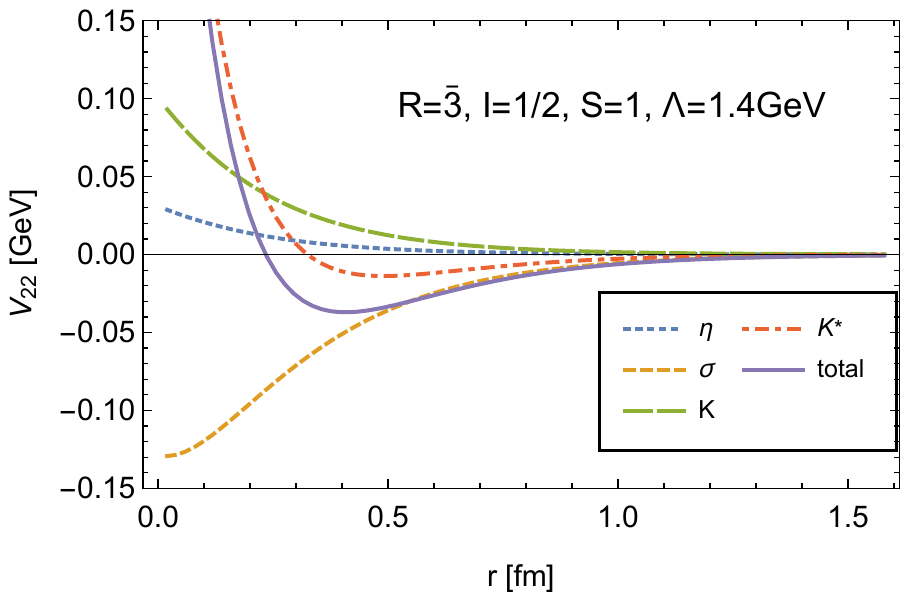}\\
\end{tabular}
\caption{The interaction potentials for the systems of the
$\bar{3}$-representation ($S=1$). $V_{11}$, $V_{12}$ and $V_{22}$
denote the ${}^3S_1\leftrightarrow {}^3S_1$, ${}^3S_1\leftrightarrow
{}^3D_1$ and ${}^3D_1\leftrightarrow {}^3D_1$ transitions
potentials, respectively. The upper panel is for $(R, I) = (\bar{3},
0)$ while the lower panel is for $(R,I) = (\bar{3}, 1/2)$.
}\label{pttl3s1}
\end{figure}

Since the spins of the systems belonging to the
$\bar{3}$-representation are $1$, the ${}^3S_1$ and ${}^3D_1$
channels couple with each other. We plot the potentials for each
exchanged boson in Fig.~\ref{pttl3s1}. From the plots, one can see
clearly that for the $(R,I) = (\bar{3},0)$ case, the $\pi$- and
$\omega$-exchanges provide repulsive potential while the $\rho$- and
$\sigma$-exchanges supply the attractive force in the ${}^3S_1$
channel. The contribution of the $\eta$-exchange is almost
negligible. The total potential is attractive in the whole range. In
the ${}^3D_1$ channel, only the $\sigma$-exchange provides
considerably attractive force. As a result, the total potential is
repulsive in the short-range, less than $0.4$ fm, while weakly
attractive in the range $0.4 < r < 1.5$ fm. In the ${}^3S_1
\leftrightarrow {}^3D_1$ transition potential, the contributions of
the $\rho$- and $\pi$-exchanges cancel each other significantly. As
a result, the total potential is weakly attractive. Although the
exchanged bosons for the  $(R, I) = (\bar{3}, 1/2)$ case are
different from those for the $(R, I) = (\bar{3}, 0)$ case, the total
potentials for both of the two cases are very similar, see
Fig.~\ref{pttl3s1}.

\begin{figure}[htp]
\centering
\begin{tabular}{ccc}
\includegraphics[width=0.3\textwidth]{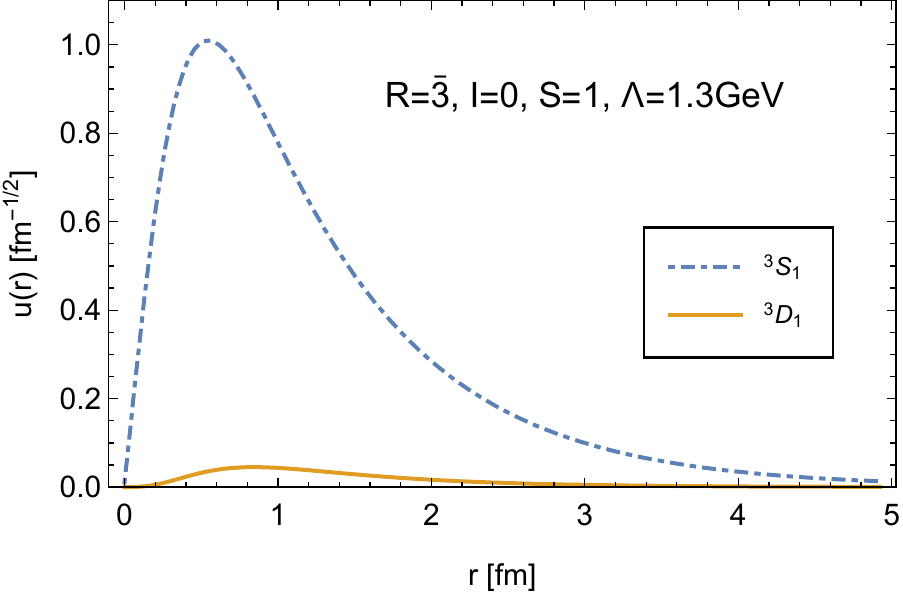}~&~
\includegraphics[width=0.3\textwidth]{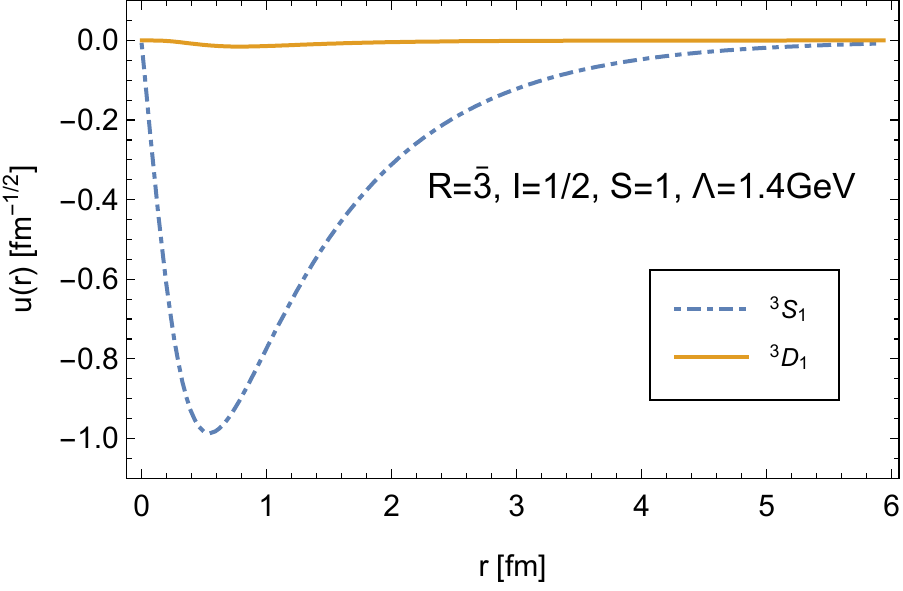}~&~
\includegraphics[width=0.3\textwidth]{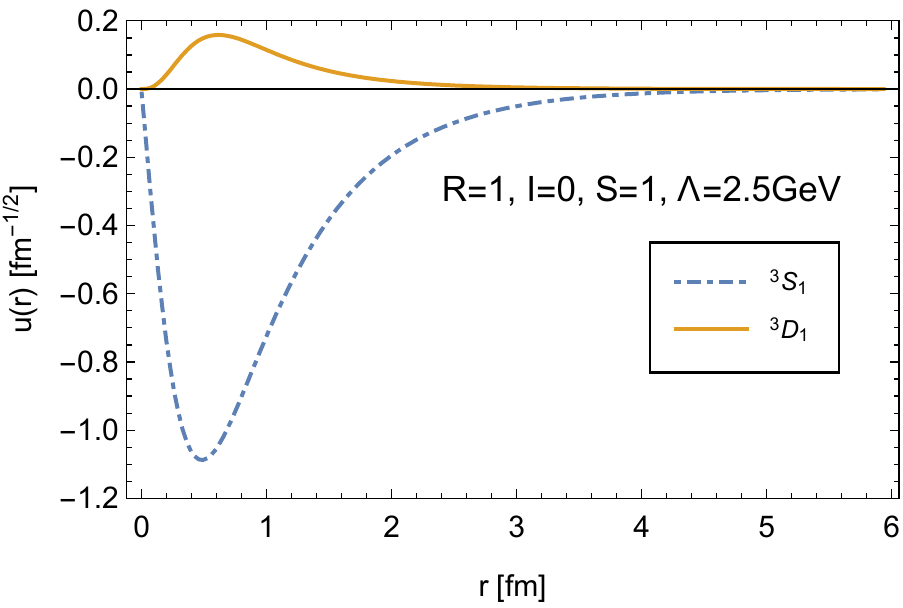}\\
\includegraphics[width=0.3\textwidth]{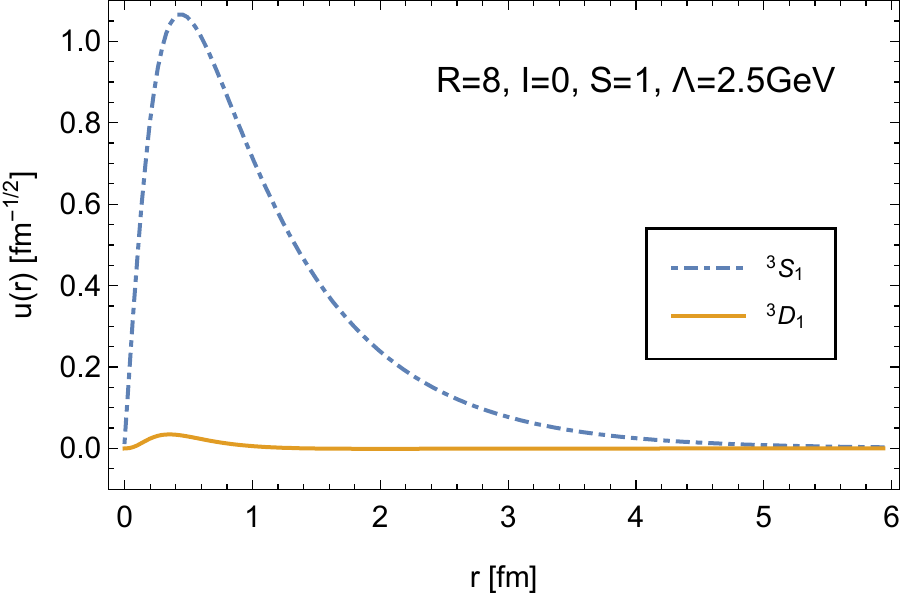}~&~
\includegraphics[width=0.3\textwidth]{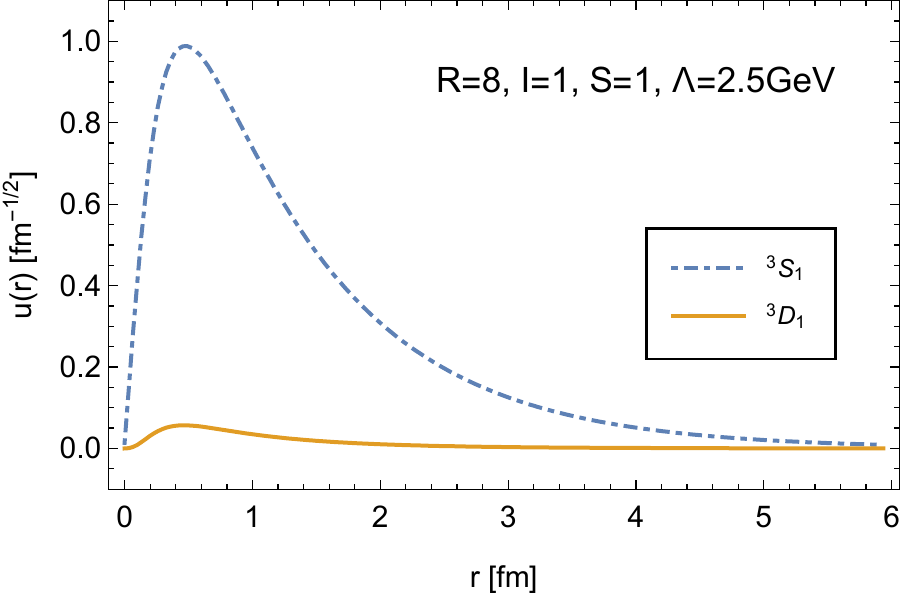}~&~
\includegraphics[width=0.3\textwidth]{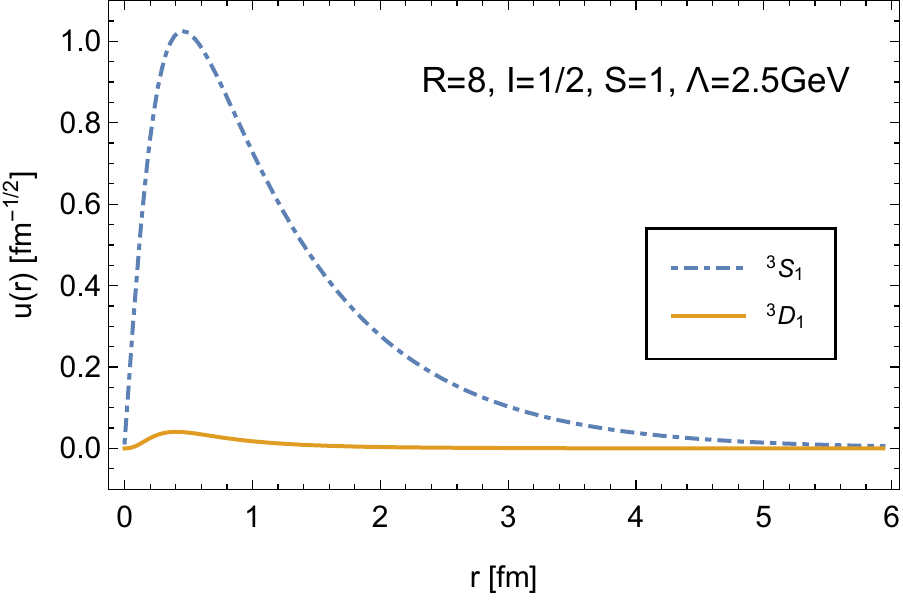}\\
\end{tabular}
\caption{(Color online).The radial wave functions $u(r) = y(r)r$
for the spin-triplet channels. }\label{wavefunc}
\end{figure}

The numerical results for systems $(R, I) = (\bar{3}, 0)$ and
$(\bar{3}, 1/2)$ are given in Table~\ref{Table:B-BS=1}. Although
the results depend on the cutoff, one can see
clearly that for both of the two systems belonging to the
$\bar{3}$-representation, there exist loosely bound states with
binding energies around a few MeV for a reasonable cutoff around
$1.2$ GeV. To investigate the effect of the short-range interaction
in forming the bound states, we also present the results without the
contact delta interaction. We find that the binding energy almost
doubles for the same cutoff once the delta interaction is switched
off since the contact interaction is repulsive. But the qualitative
features do not change very much. We also notice that the
probability of the $D$ wave is tiny, less than $0.4\%$. This is not
surprising since the potential for the transition
${}^3S_1\leftrightarrow{}^3D_1$ is very weak. The radial wave
function $u(r) = y(r)r$ for the individual channel is shown in
Fig.~\ref{wavefunc}. We conclude that the systems of the
$\bar{3}$-representation are good candidates of the deuteron-like
states.

\begin{table}
  \centering
 \caption{The binding solutions for the $B_{cc}B_{cc}$ systems.
``$\Lambda$" is the cutoff parameter. ``B.E." means the binding
energy while $r_{rms}$ is the root-mean-square radius. $P_S$ is the
probability (\%) of the $S$ wave. }\label{Table:B-BS=1}
\begin{tabular}{lcccccccc}
\toprule[1pt]\toprule[1pt]
\multicolumn{4}{c}{With contact term}& \multicolumn{4}{c}{Without contact term}\\
Systems & $\Lambda$ (GeV) & B.E (Mev) & $r_{rms}$ (fm) & $P_{S}$
(\%) & $\Lambda$ (GeV) & B.E (Mev) & $r_{rms}$ (fm) & $P_{S}$ (\%)
\tabularnewline \midrule[1pt] $(\bar{3},\frac{1}{2})$              &
1.2 & 0.56  & 3.45 & 99.98 & 1.2 & 2.41 & 1.85 & 99.96
\tabularnewline
                                     & 1.5 & 17.76 & 0.86 & 99.99 & 1.5 & 34.55& 0.66 & 99.99
 \tabularnewline
                                     & 1.9 & 60.58 & 0.55 & 99.94 & 1.9 &116.04& 0.42 & 99.93
 \tabularnewline
$(\bar{3},0)$                        & 1.1 & 0.68  & 3.23 & 99.74 &
1.1 & 3.28 & 1.66 & 99.69 \tabularnewline
                                     & 1.3 & 12.25 & 1.01 & 99.79 & 1.3 & 25.07 & 0.77 & 99.86
 \tabularnewline
                                     & 1.5 & 33.20 & 0.70 & 99.93 & 1.5 & 61.46 & 0.55 & 99.97
 \tabularnewline
\bottomrule[1pt]\bottomrule[1pt]
\end{tabular}

\end{table}

\subsubsection{$6$-representation, S = 0}

\begin{figure}[htp]
\centering
\begin{tabular}{ccc}
\includegraphics[width=0.30\textwidth]{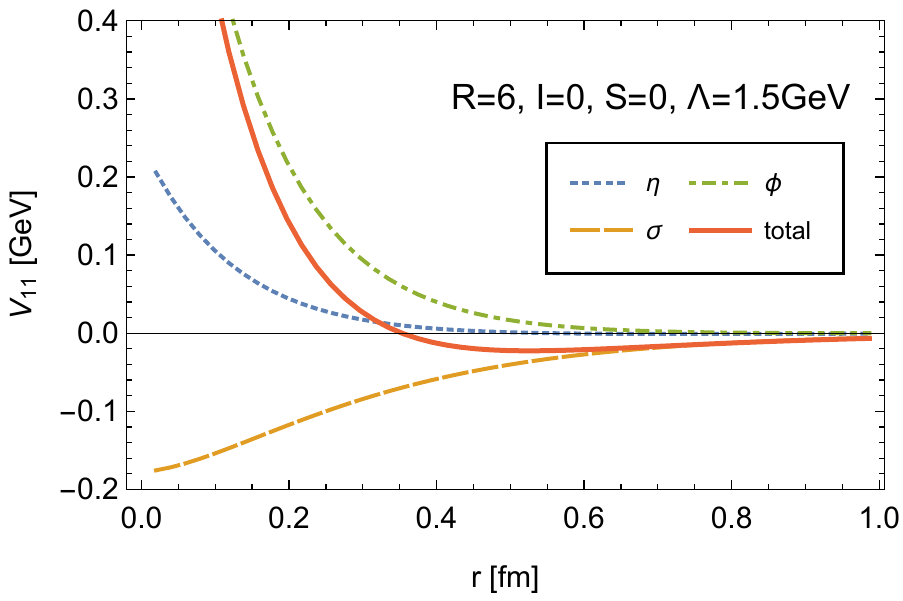}~&~
\includegraphics[width=0.30\textwidth]{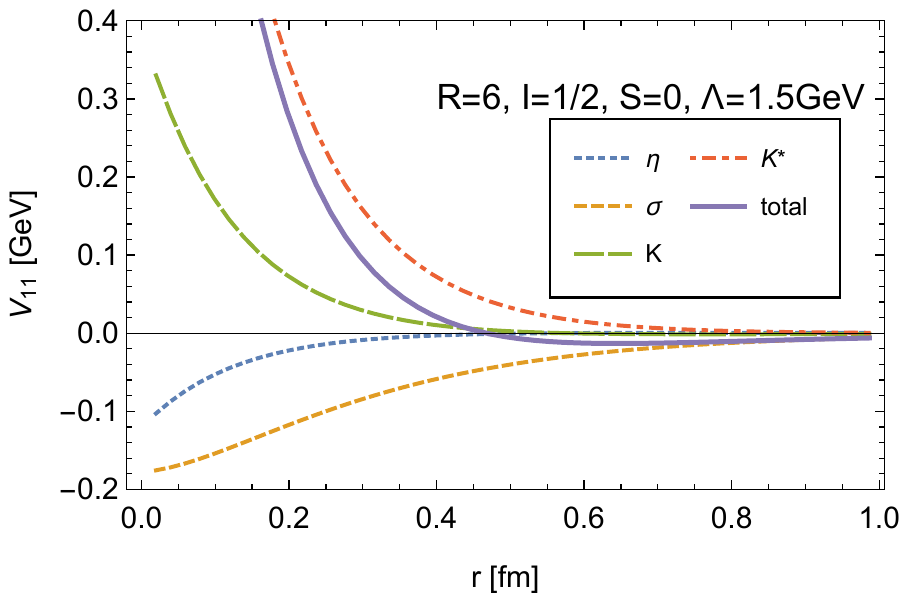}~&~
\includegraphics[width=0.30\textwidth]{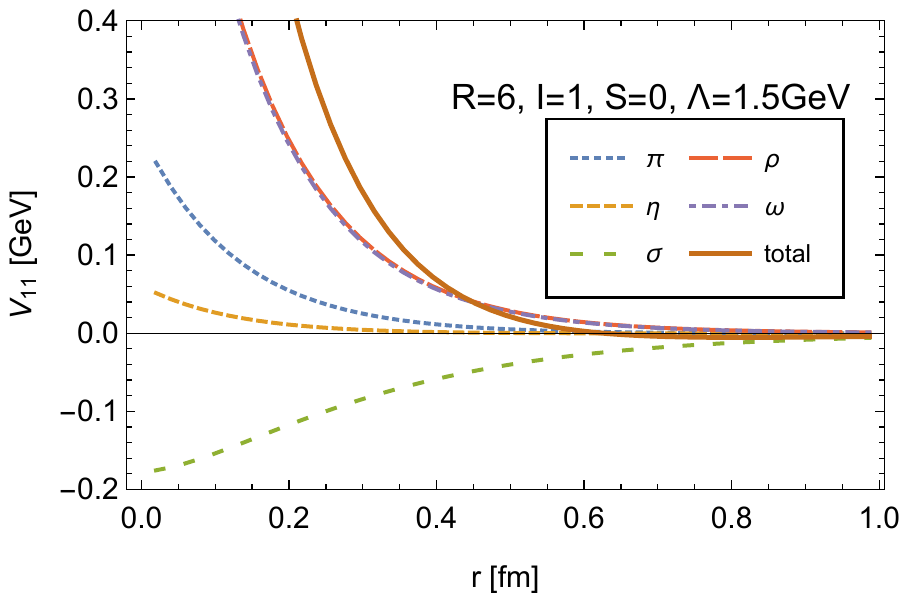}\\
\end{tabular}
\caption{(Color online). The interaction potentials for the systems
of the $6$-representation. }\label{pttl6s0}
\end{figure}

The systems of the $6$-representation  are simpler since they are
all spin-singlets. We show the potential for each boson-exchange in
Fig.~\ref{pttl6s0}. From the plots, one can see clearly that the
total potentials for all of the three systems are repulsive in the
range, less than $0.4$ fm, for the cutoff around $1.5$ GeV. The
numerical results are given in Table~\ref{Table:B-BS=0}. For the
system $(R,I)=(6,1)$, we fail to obtain any binding solutions. For
the systems $(R,I) = (6,1/2)$ and $(6, 0)$, we could not obtain
binding solutions until we increase the cutoff to be $5.4$ GeV and
$3.8$ GeV respectively. If we switch off the contact delta
interaction, a loosely bound state is obtained for $(6,1)$ with
$\Lambda = 1.9$ GeV, for $(6,1/2)$ with $\Lambda = 1.6$ GeV and for
$(6,0)$ with $\Lambda = 1.5$ GeV. However, the contact delta
interaction in the spin-0 systems with $6$-representation is
strongly repulsive. Moreover, Pauli principle may forbid the four
charm quarks at the origin simultaneously. Therefore, we conclude
that there do not exist the molecular states for the systems of the
$6$-representation.

\begin{table}
\centering
\caption{The binding solutions for the systems of the
$6$-representation.  ``$\times$" means that no binding solutions are
obtained. }\label{Table:B-BS=0}
\begin{tabular}{lcccccc}
\toprule[1pt]\toprule[1pt]
 &\multicolumn{3}{c}{With contact term}& \multicolumn{3}{c}{Without contact term}\\
Systems & $\Lambda$ (GeV) & B.E (MeV) & $r_{rms}$ (fm)  & $\Lambda$
(GeV) & B.E (Mev) & $r_{rms}$ (fm) \tabularnewline \midrule[1pt]
$(6,1)$ & $\times$ & $\times$ & $\times$            & 1.9 & 0.31 &
4.27 \tabularnewline
                   &  &  &                          & 3.0 & 3.43 & 1.56
 \tabularnewline
                   &  &  &                          & 3.6 & 5.11 & 1.31
 \tabularnewline
$(6,\frac{1}{2})$            & 5.4 & 0.14 & 5.25 &  1.6 & 4.69 &
1.40 \tabularnewline
                              & 6.6 & 1.29 & 2.45 &  1.9 & 12.38 & 0.95
 \tabularnewline
                              & 7.5 & 2.65 & 1.80 &  2.5 & 31.10 & 0.65
 \tabularnewline
$(6,0) $                      & 3.8 & 0.10 & 5.55 &  1.5 & 5.50 &
1.31 \tabularnewline
                              & 4.5 & 1.27 & 2.48 &  1.7 & 14.80 & 0.89
 \tabularnewline
                              & 5.0 & 2.74 & 1.80 &  2.0 & 34.16 & 0.64
 \tabularnewline
\bottomrule[1pt]\bottomrule[1pt]
\end{tabular}
  \end{table}

\subsection{$B_{cc}\bar{B}_{cc}$ systems }

For the baryon-antibaryon systems, there is no constraint from Pauli
Principle. All the systems can be both spin-singlet ($S=0$) and
spin-triplet ($S=1$). We present the results according to the spin
of the system, i.e., spin-singlet and spin-triplet. The $S=0$ and $S=1$ potentials are shown in Fig.~\ref{pttl81s0} and \ref{pttl81s1}, respectively.

\subsubsection{$B_{cc}\bar{B}_{cc}$, spin-singlet}

\begin{figure}[htp]
\centering
\begin{tabular}{ccc}
\includegraphics[width=0.35\textwidth]{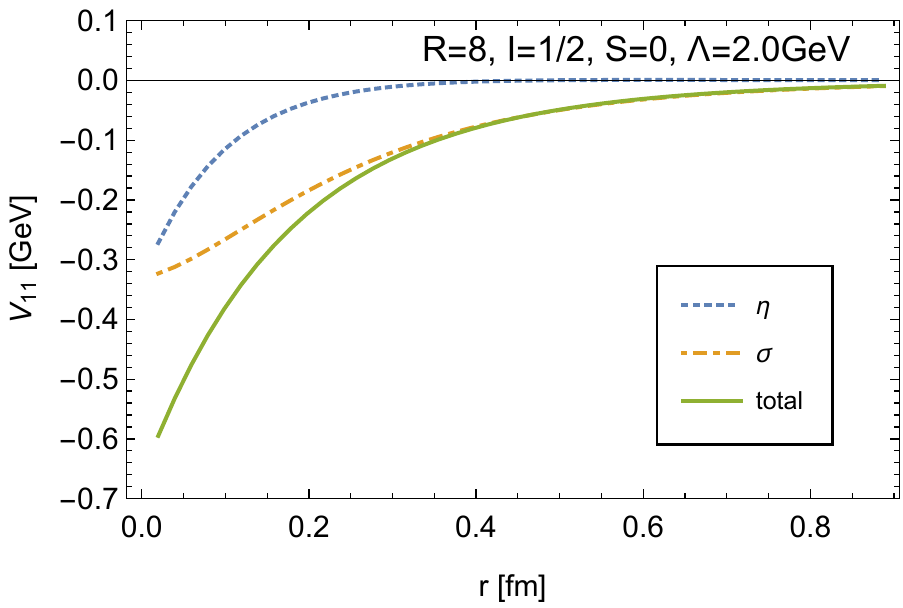}~&~
\includegraphics[width=0.35\textwidth]{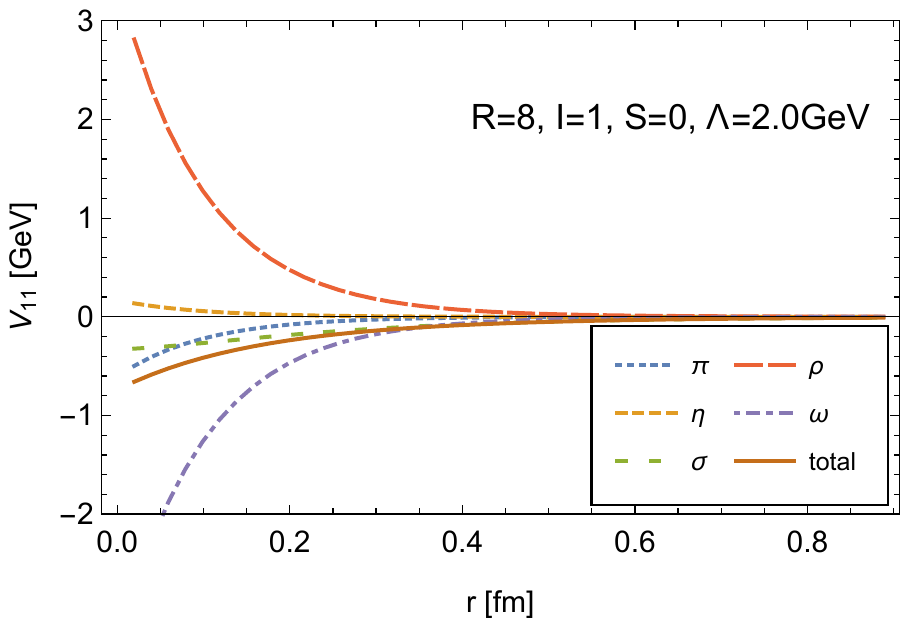}\\
\includegraphics[width=0.35\textwidth]{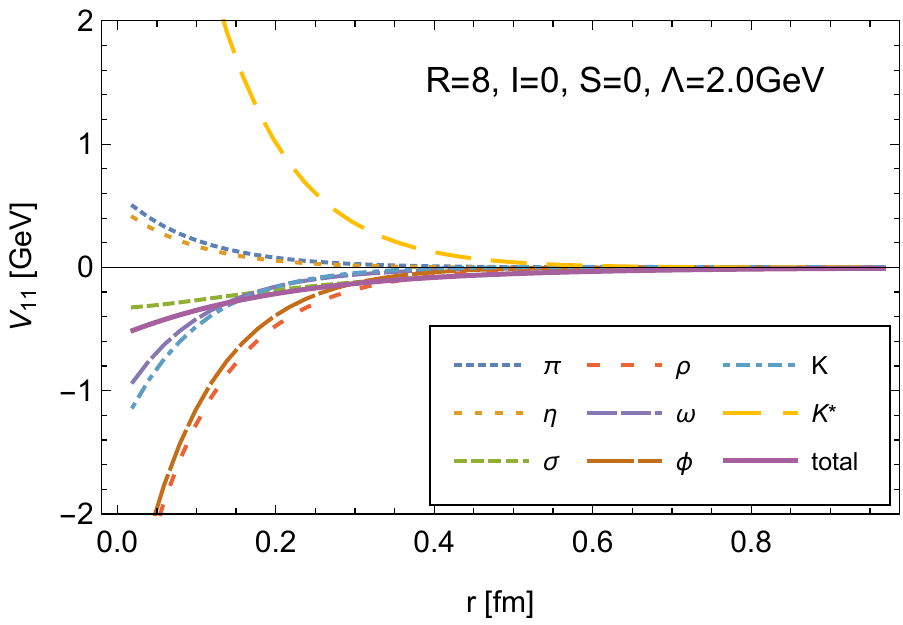}~&~
\includegraphics[width=0.35\textwidth]{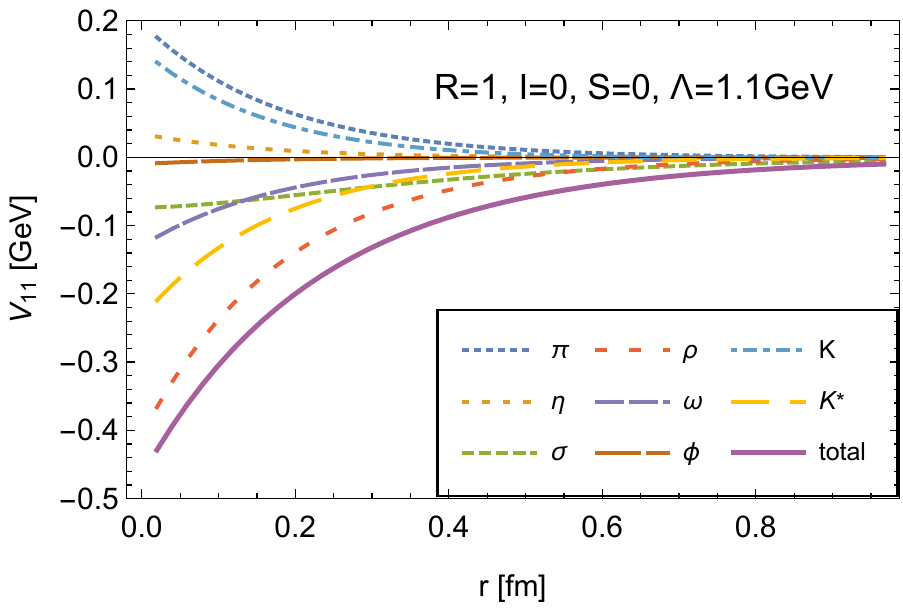}\\
\end{tabular}
\caption{(Color online). The interaction potentials of the
spin-singlet $B_{cc}\bar{B}_{cc}$ systems.}\label{pttl81s0}
\end{figure}

\begin{figure}[htp]
\centering
\begin{tabular}{ccc}
\includegraphics[width=0.3\textwidth]{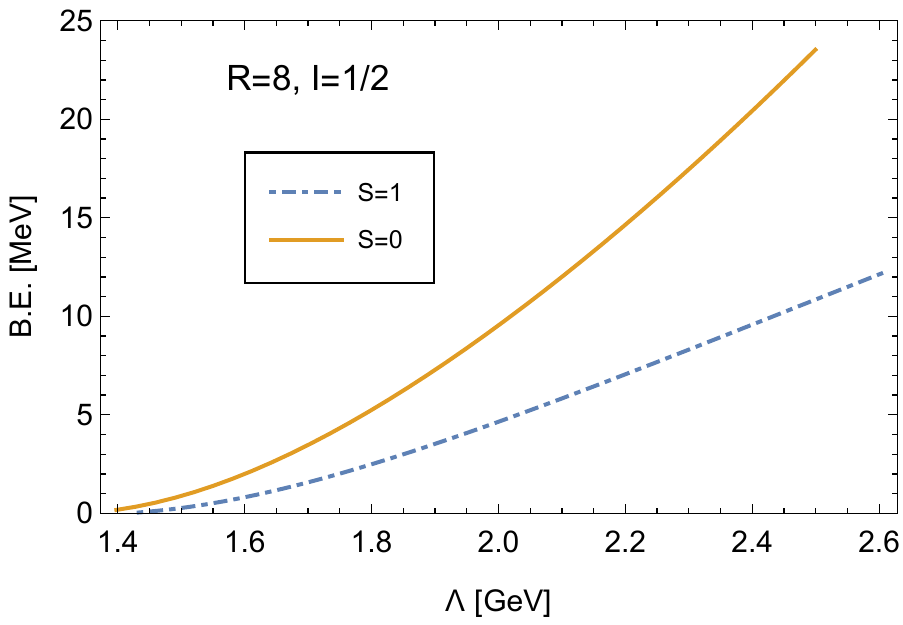}~&~
\includegraphics[width=0.3\textwidth]{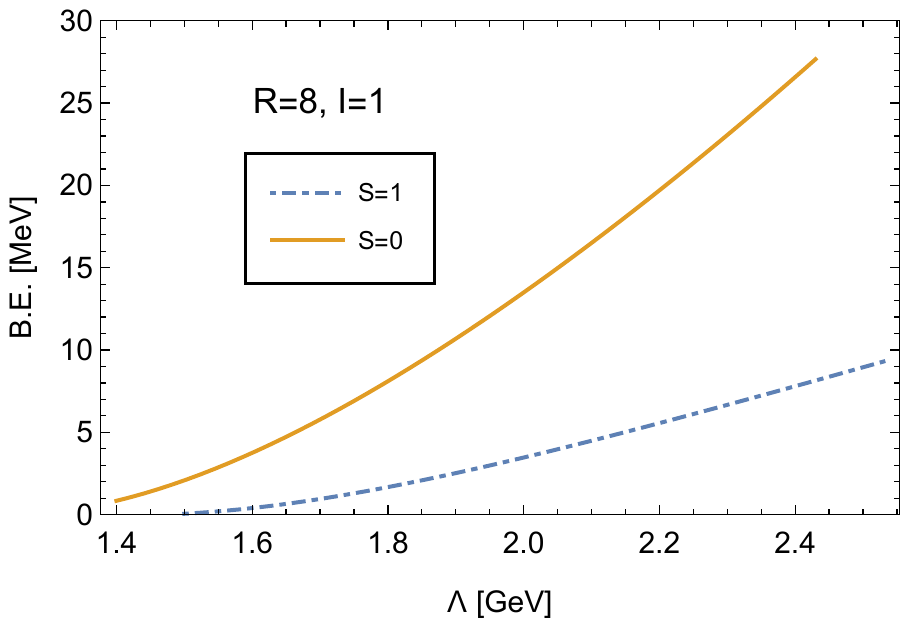}~&~
\includegraphics[width=0.3\textwidth]{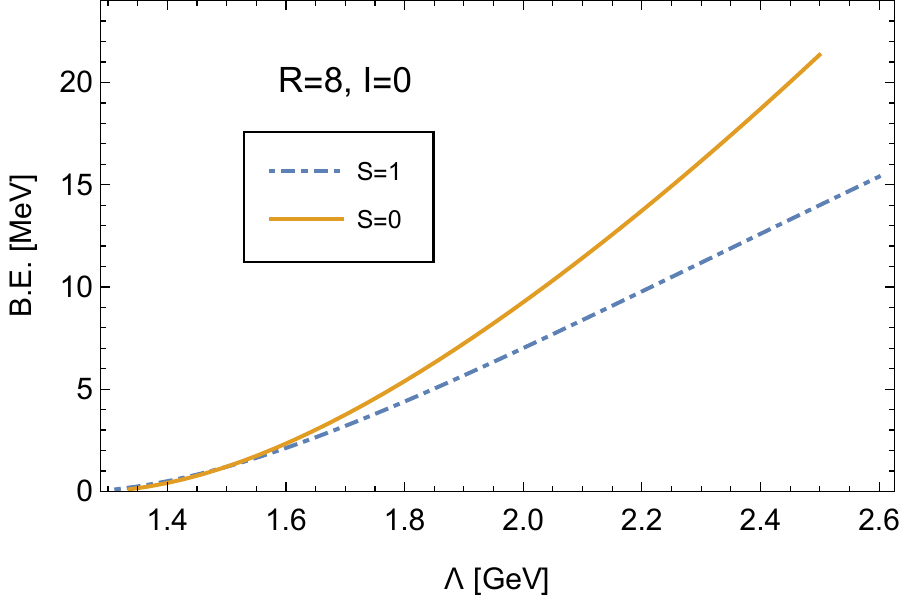}\\
\includegraphics[width=0.3\textwidth]{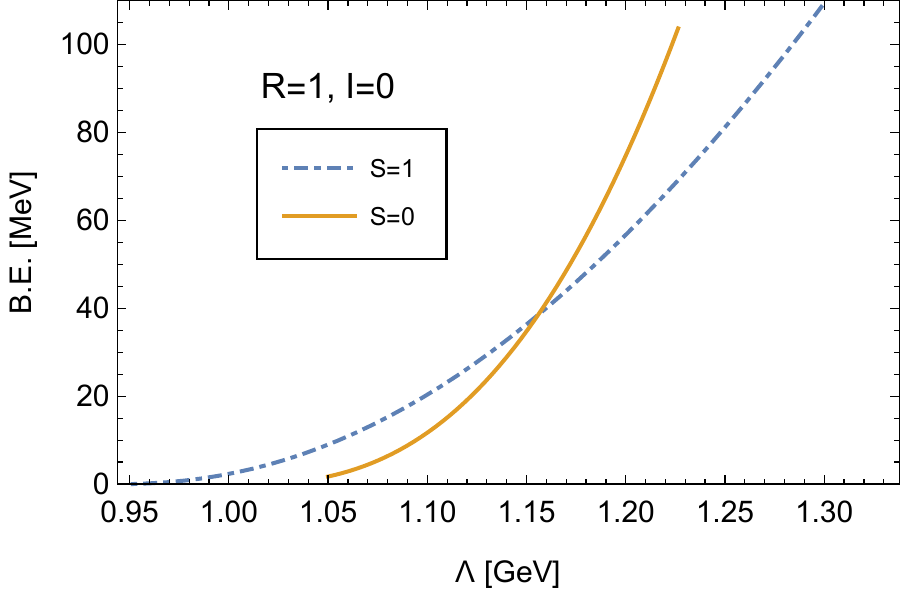}~&~
\includegraphics[width=0.3\textwidth]{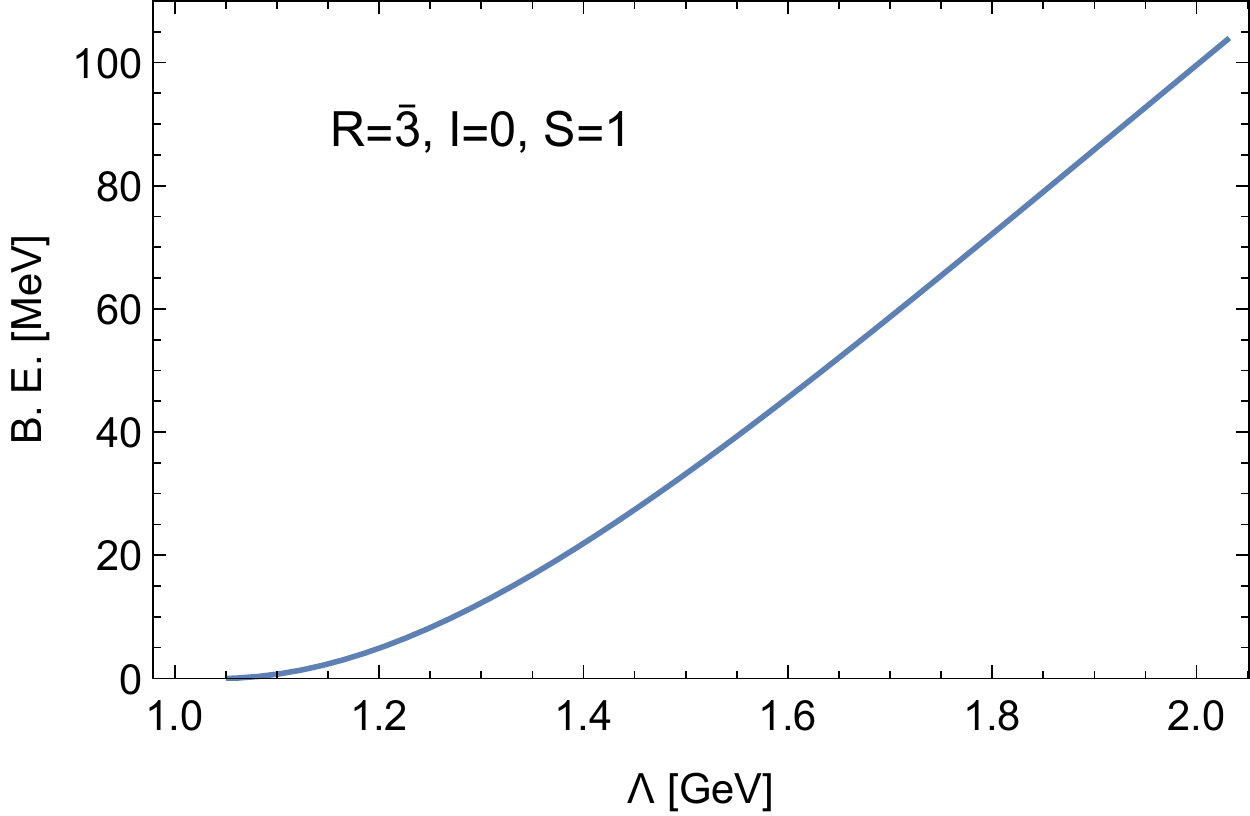}~&~
\includegraphics[width=0.3\textwidth]{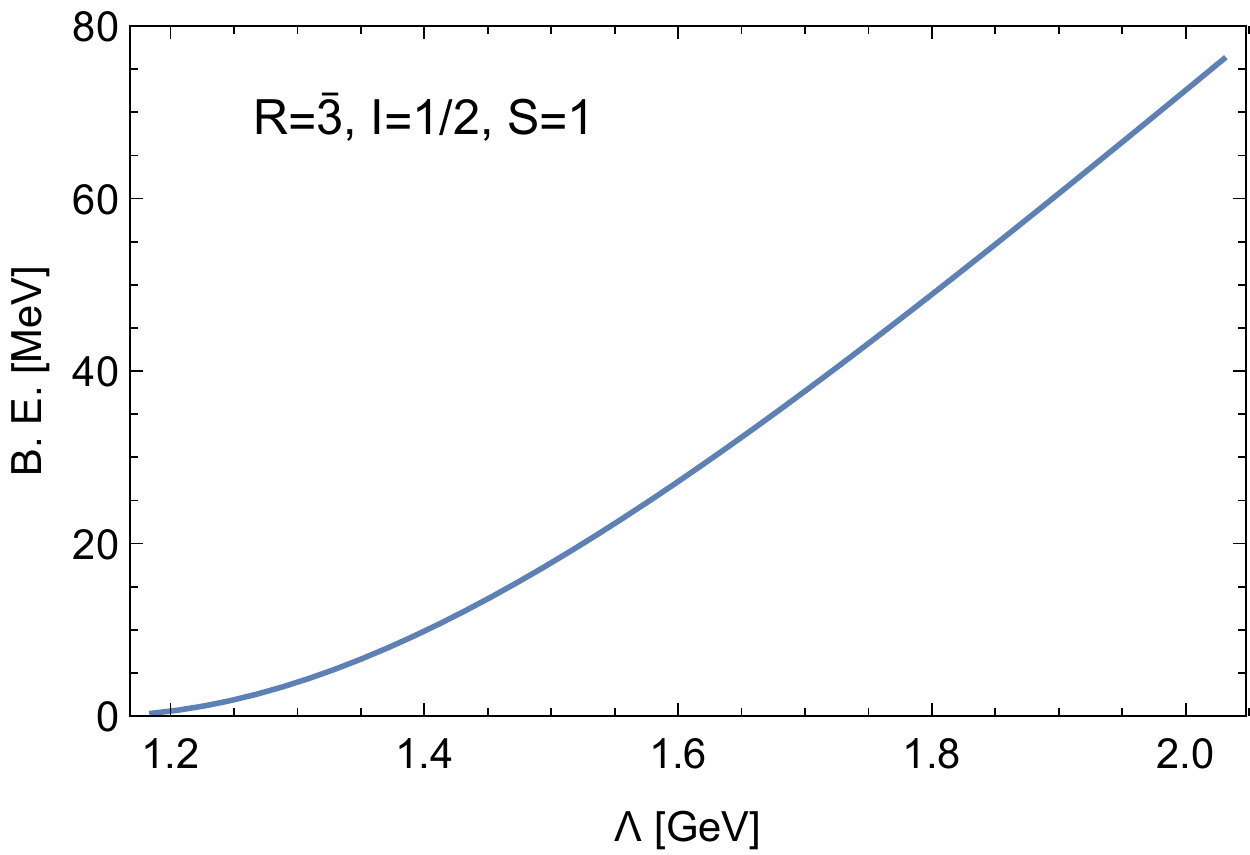}\\
\end{tabular}
\caption{(Color online). The binding energy versus the cutoff
parameter. The contact interaction is included. }\label{cut-be}
\end{figure}

For the system $(R,I)=(8,1/2)$, only $\eta$- and $\sigma$-exchanges
are allowed while all the $\eta$-, $\sigma$-, $\pi$-, $\rho$- and
$\omega$- exchanges contribute to the system $(8,1)$. For the system
$(8,0)$ and $(1,0)$, additional $K$-, $K^*$- and $\phi$-exchanges
are also allowed. We give the numerical binding-solution results in Table~\ref{Table:B-barBS=0}.
Interestingly, we obtain a loosely bound state for the system
$(R,I)=(8,1/2)$ for the cutoff in the range $1.5 < \Lambda < 2.0$
GeV, both with and without the contact interaction. For this bound
state, both the $\eta$- and $\sigma$-exchanges supply attractive
force, see Fig.~\ref{pttl81s0}. From Fig.~\ref{cut-be}, one can also
see that the binding solutions depend weakly on the cutoff
parameter, which indicates the system $(R,I)=(8,1/2)$ is a good
candidate of the molecular state.

There also exist loosely bound states for the systems $(R,I) =
(8,1)$ and $(8,0)$, both with and without the contact interaction
for the cutoff in the range $1.5 - 2.0$ GeV. The binding energies
are a few MeV and the root-mean-square radii are both around $1$ fm.
For the system $(8, 1)$, the contributions of the $\rho$- and
$\omega$-exchanges cancel each other significantly. Both of the
$\sigma$- and $\pi$-exchanges provide attractive force while the
$\eta$-exchange supply the repulsive force. For the system $(8, 0)$,
the potential from the $K^*$-exchange is strongly repulsive. The
$\eta$- and $\pi$-exchanges also provide repulsive force while the
potentials from the $\rho$-, $\omega$-, $\sigma$-, $\phi$- and $K$-
exchanges are attractive, see Fig.~\ref{pttl81s0}. These two
interesting states are also good candidates of the molecular states.

Although we obtain binding solutions for the system $(1,0)$, the
results depend strongly on the cutoff parameter. After removing the
contact interaction, a loosely bound state is obtained for the
cutoff around $1.1 < \Lambda < 1.6$ GeV.  This system might be a
molecule candidate.

From Table \ref{Table:B-barBS=0}, one can see that the binding is larger when the contact interaction is included. 
The contact interactions of the $\pi$, $\rho$ and $\sigma$ exchanges (the isospin factor is set to 1) for the spin-singlet system are shown in Fig.~\ref{pttldeltas0}. One can see clearly that the contribution of the  $\pi$ and $\rho$ exchanges to the contact interaction are roughly equal, and both are repulsive. The $\sigma$ exchange contribution is negligible. From Table \ref{Table:Isospinfac},
the summation of the isospin factors of the vector mesons for
8-representation systems are 0. Thus, the vector meson exchange
contribution to the contact interaction almost cancels out. The
attractive contact interaction mainly arise from the pseudoscalar
exchanges. For the 1-representation system, the attractive contact
interaction is the result of the cancellation of the vector meson
exchanges with the pseudoscalar exchanges.

\begin{table}
  \centering
 \caption{The binding solutions of the spin-singlet
$B_{cc}\bar{B}_{cc}$ systems.}\label{Table:B-barBS=0}
\begin{tabular}{lccccccc}
\toprule[1pt]\toprule[1pt]
 &\multicolumn{3}{c}{With contact term}& \multicolumn{3}{c}{Without contact term}\\
Systems & $\Lambda$ (GeV) & B.E. (MeV) & $r_{rms}$ (fm) & $\Lambda$
(GeV) & B.E. (MeV) & $r_{rms}$ (fm) \tabularnewline \midrule[1pt]
$(8,1)$   & 1.3 & 0.11 & 5.41 & 1.5 & 0.26 & 4.49\tabularnewline
          & 1.6 & 3.75 & 1.50 & 1.6 & 0.83 & 2.87\tabularnewline
          & 2.0 & 13.49 & 0.89 &1.9 & 3.77 & 1.50\tabularnewline
          & 2.5 & 30.25 & 0.64 &2.6 & 13.28 & 0.89\tabularnewline
$(8,\frac{1}{2})$      & 1.4 & 0.18 & 4.93 & 1.5 & 0.05 &
5.96\tabularnewline
                       & 1.6 & 2.00 & 1.96 & 1.8 & 1.70 & 2.11\tabularnewline
                       & 2.0 & 9.54 & 1.02 & 2.0 & 3.53 & 1.54\tabularnewline
                       & 2.5 & 23.55 & 0.70 & 2.5 & 9.10 & 1.04\tabularnewline
$(8,0)$   & 1.4 & 0.42 & 3.84 & 1.4 & 0.04 & 6.06\tabularnewline
          & 1.6 & 2.34 & 1.85 & 1.6 & 1.08 & 2.59\tabularnewline
          & 2.0 & 9.25 & 1.04 & 2.0 & 4.96 & 1.35\tabularnewline
          & 2.5 & 21.36 & 0.74 & 2.5 & 10.63 & 0.98\tabularnewline
$(1,0)$   &1.05 & 1.71  & 2.17 & 1.1 & 0.08 & 5.71\tabularnewline
          & 1.1 & 11.68 & 0.99 & 1.2 & 1.00 & 2.74\tabularnewline
          & 1.2 & 74.73 & 0.48 & 1.3 & 2.58 & 1.83\tabularnewline
          & 1.3 & 216.46 & 0.32 & 1.6 & 9.40 & 1.09\tabularnewline
\bottomrule[1pt]\bottomrule[1pt]
\end{tabular}
\end{table}

\subsubsection{$B_{cc}\bar{B}_{cc}$, spin-triplet}

\begin{figure}[htp]
\centering
\begin{tabular}{ccc}
\includegraphics[width=0.30\textwidth]{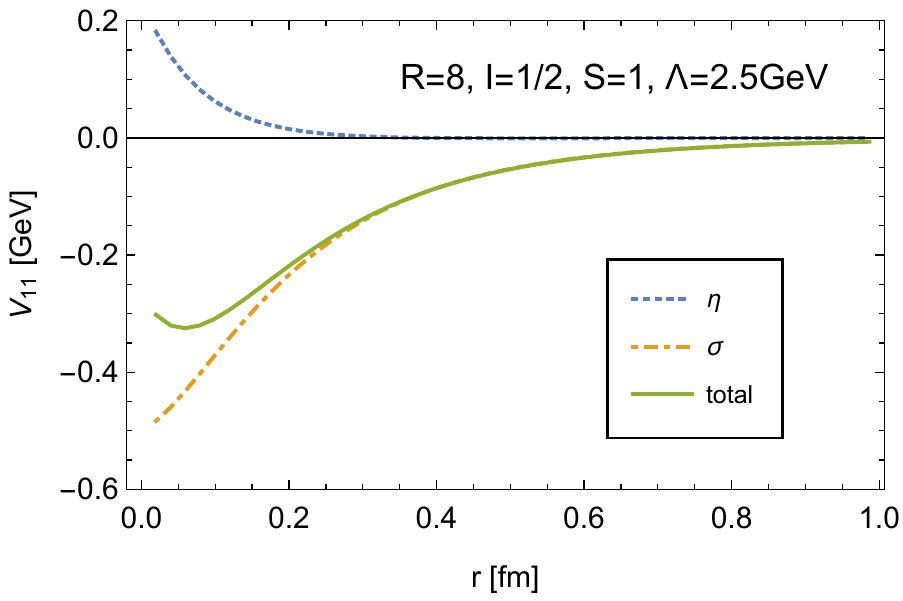}~&~
\includegraphics[width=0.30\textwidth]{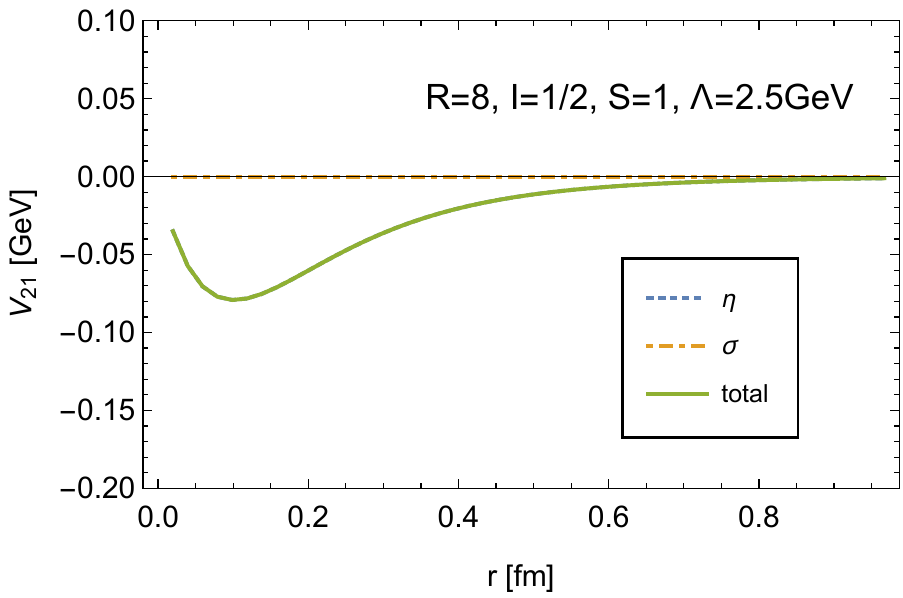}~&~
\includegraphics[width=0.30\textwidth]{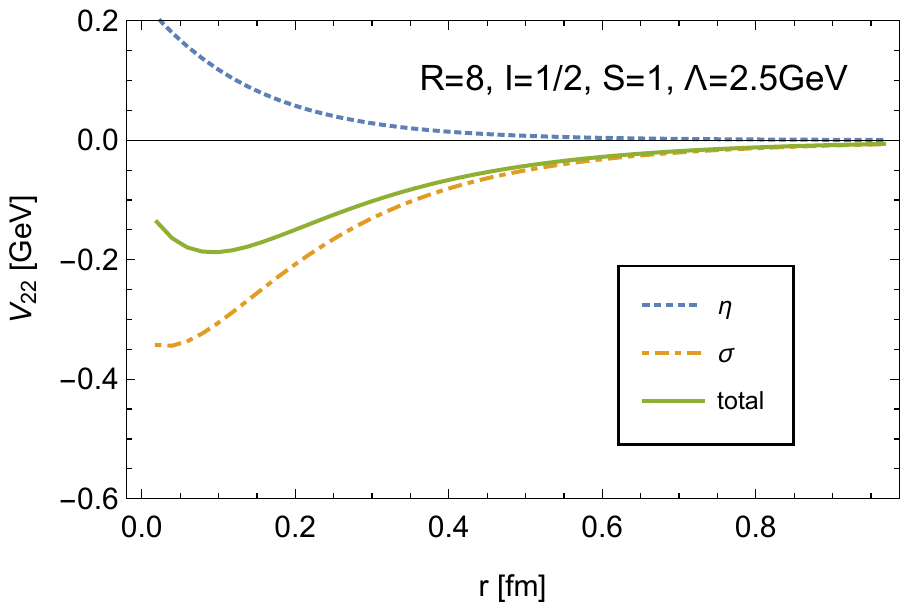}\\
\includegraphics[width=0.30\textwidth]{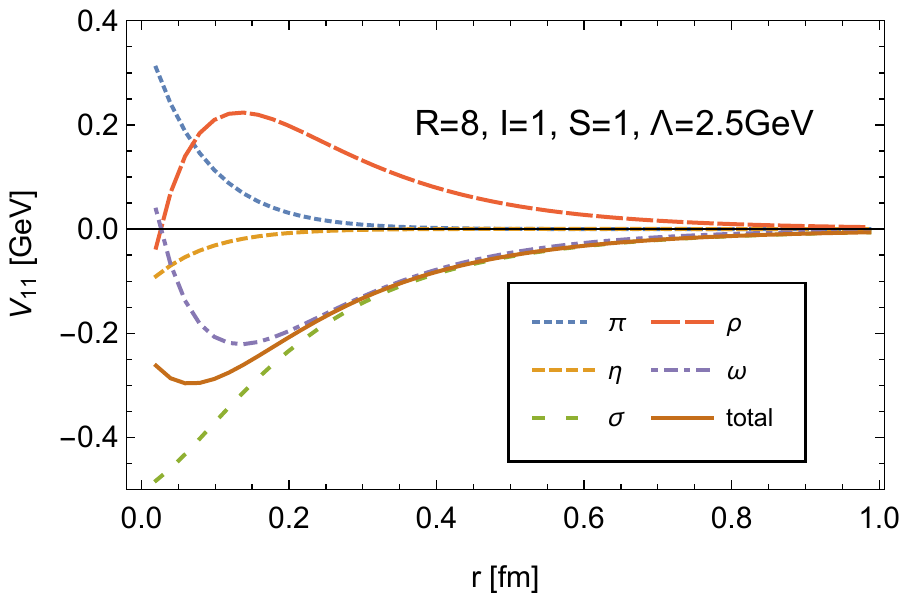}~&~
\includegraphics[width=0.30\textwidth]{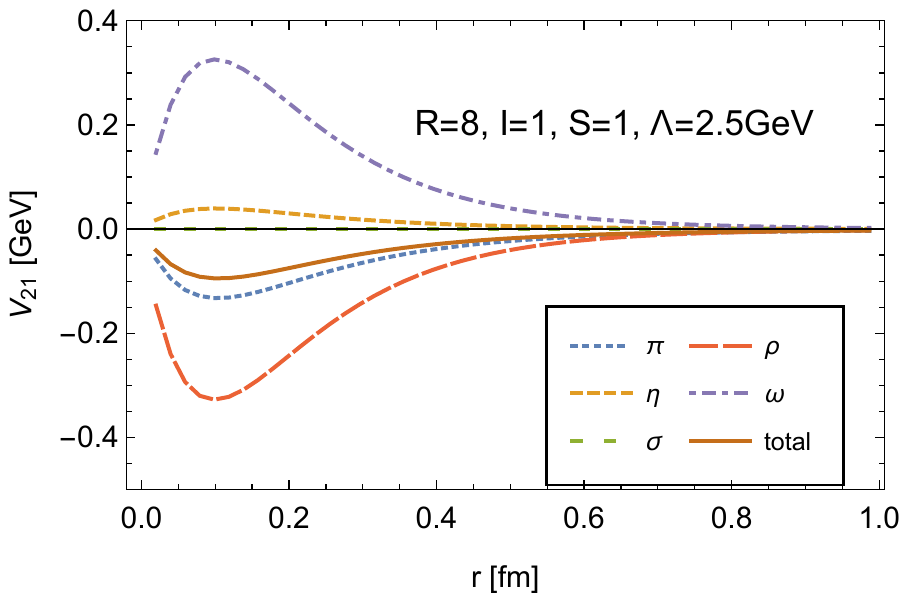}~&~
\includegraphics[width=0.30\textwidth]{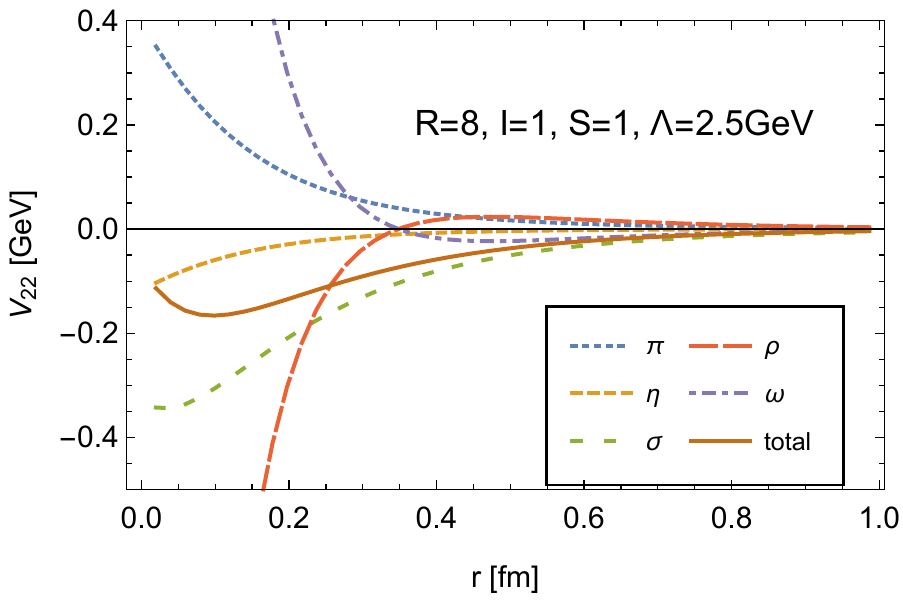}\\
\includegraphics[width=0.30\textwidth]{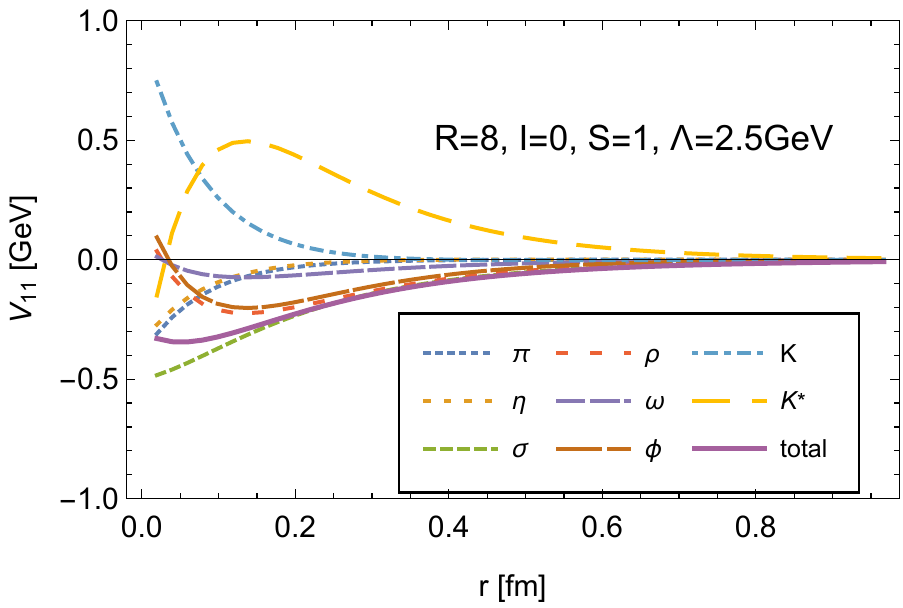}~&~
\includegraphics[width=0.30\textwidth]{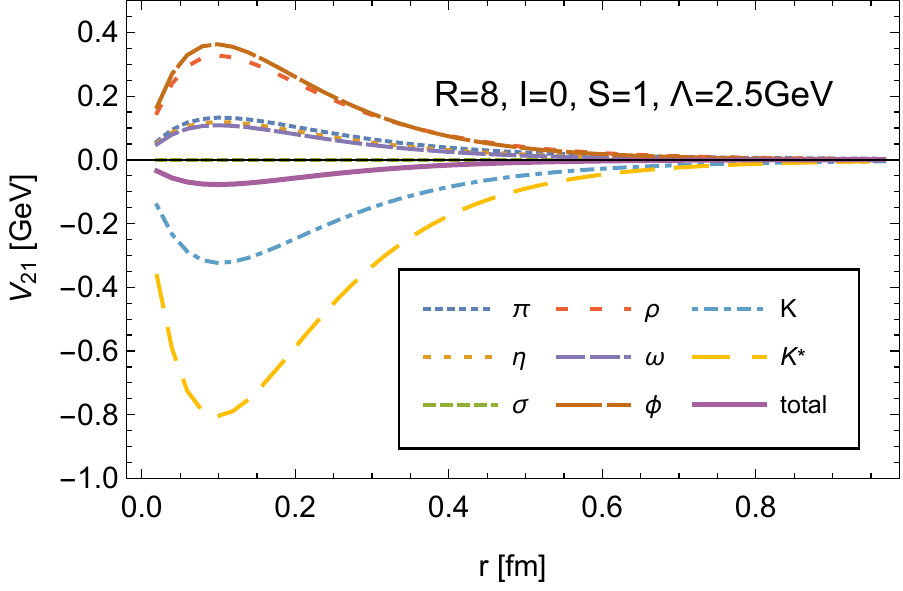}~&~
\includegraphics[width=0.30\textwidth]{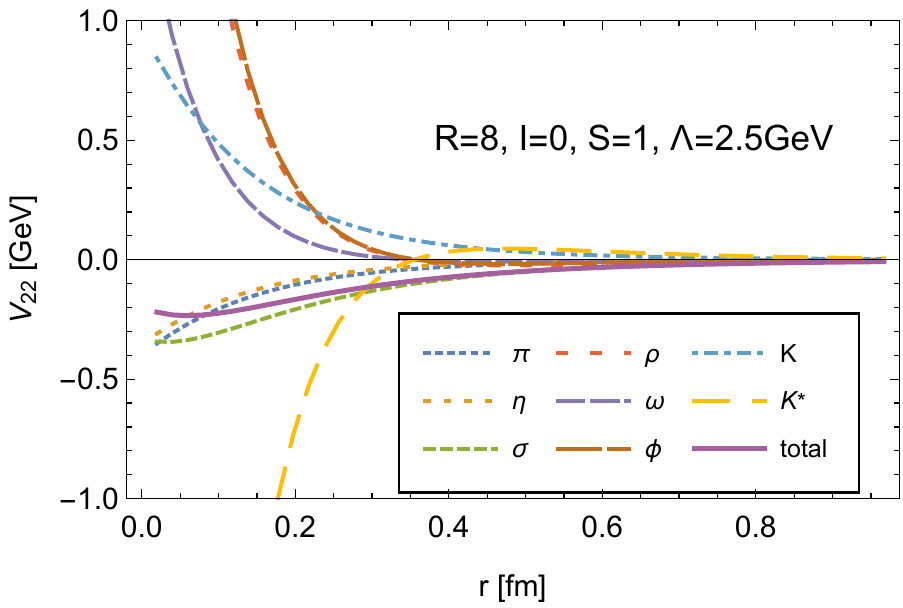}\\
\includegraphics[width=0.30\textwidth]{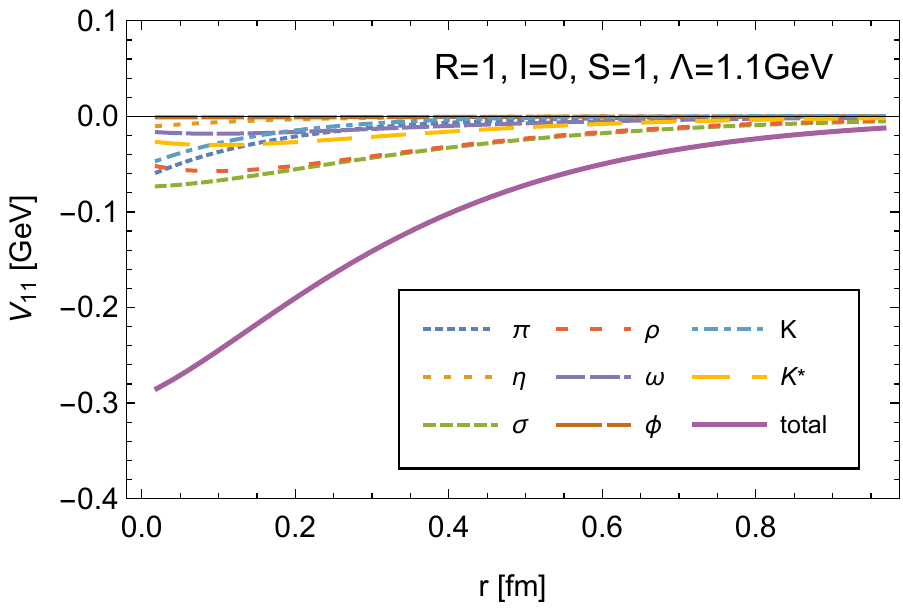}~&~
\includegraphics[width=0.30\textwidth]{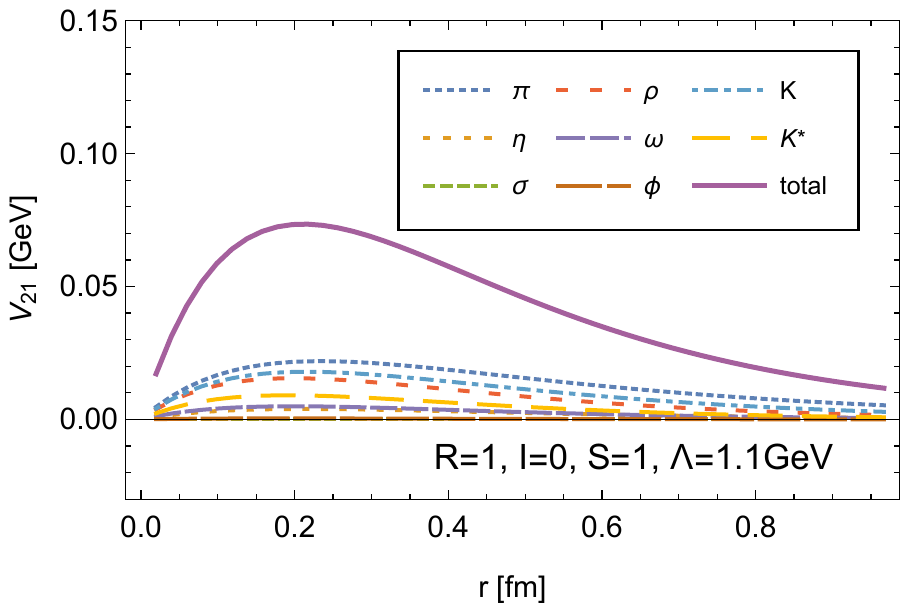}~&~
\includegraphics[width=0.30\textwidth]{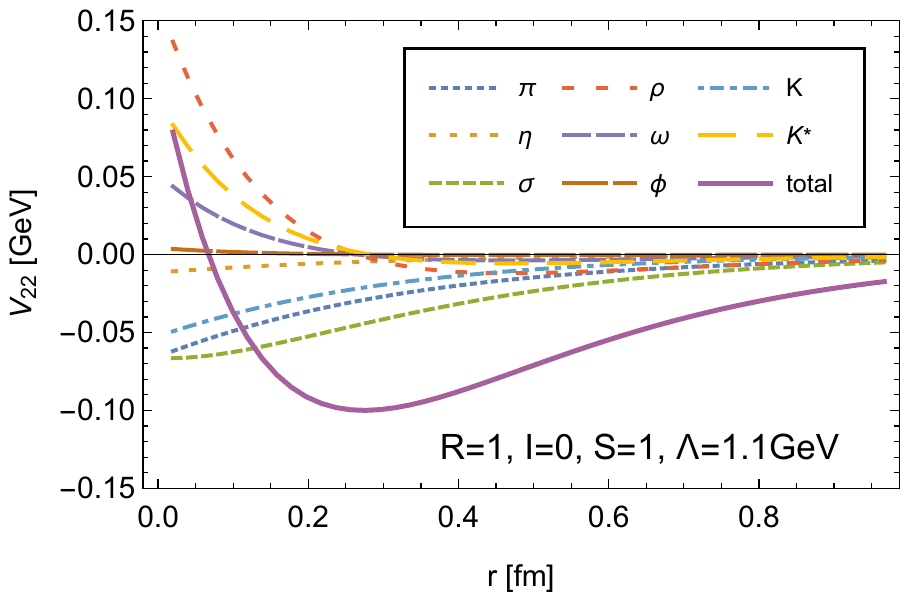}\\
\end{tabular}
\caption{(Color online). The interaction potentials of the
spin-triplet $B_{cc}\bar{B}_{cc}$ systems.}\label{pttl81s1}
\end{figure}

For the spin-triplet case, we show the potentials in
Fig.~\ref{pttl81s1} and present the binding solutions in
Table~\ref{Table:B-barBs=1}. Similar to the spin-singlet case, we
also obtain loosely bound states for a reasonable cutoff in the
spin-triplet sector. These states are very interesting and are good
candidates of the molecular states. For example, we obtain a loosely
bound state for the system $(R,I)=(8,1)$ which has binding energy
$0.05 - 3.47$ MeV and root-mean-square radius $5.98 - 1.56$ fm for
the cutoff around $1.5 - 2.0$ GeV.  With the same cutoff, a loosely
bound state of the system $(8, 1/2)$ with binding energy $0.27 -
4.65$ MeV and root-mean-square radius $4.45 - 1.39$ fm is obtained.
Similarly, for the system $(8, 0)$, we obtain a loosely bound state
with binding energy $0.06 - 7.01$ MeV for the cutoff around $1.3 -
2.0$ GeV. All these three states $(R,I) = (8, 1)$, $(8,1/2)$, and
$(8,0)$ are good candidates of the molecular states. We also obtain
binding solutions for the system of the $1$-representation $(1,0)$.
Unfortunately, the results depend strongly on the cutoff.

Very interestingly, we also find that for the spin-triplet case the
results change very little by removing the contact interaction. This
means that the contact interaction plays a minor role in the
formation of the bound states in the spin-triplet sector. The
contribution of the $D$-wave for the systems belonging to the
8-representation is less than $0.4 \%$, similar to that in the
baryon-baryon case. In contrast, the $D$-wave plays a more important
role in the 1-representation system for $\Lambda=1.1$ GeV.

Compared with the spin-singlet systems, the spin-triplet systems have a weaker dependence on the contact interaction. For the S wave, the contact interaction only arise from the spin-spin interaction. And the matrix elements of the
spin-spin operator for $S=1$ is 1 while that for $S=0$ is $-3$. Thus the results for the spin-triplet systems change less by removing the contact interaction, compared with the spin-singlet systems.

\begin{table}
\centering
\caption{ The binding solutions of the spin-triplet
$B_{cc}\bar{B}_{cc}$ systems. }\label{Table:B-barBs=1}
\begin{tabular}{lcccccccc}
\toprule[1pt]\toprule[1pt]
 &\multicolumn{4}{c}{With contact term}& \multicolumn{4}{c}{Without contact term}\\
Systems & $\Lambda$ (GeV) & B.E. (MeV) & $r_{rms}$ (fm) & $P_{S}$
(\%)  & $\Lambda$ (GeV) & B.E. (MeV) & $r_{rms}$ (fm) & $P_{S}$ (\%)
\tabularnewline \midrule[1pt] $(8,1)$   & 1.5 & 0.05 & 5.98 & 99.97
& 1.5 & 0.25 & 4.56 & 99.95\tabularnewline
          & 1.6 & 0.40 & 3.91 & 99.94 & 1.6 & 0.80 & 2.95 & 99.92\tabularnewline
          & 2.0 & 3.47 & 1.56 & 99.85  & 1.9 & 3.65 & 1.53 & 99.86\tabularnewline
          & 2.5 & 8.95 & 1.06 & 99.76 &  2.3 & 8.96 & 1.05 & 99.79\tabularnewline
$(8,\frac{1}{2})$      & 1.5 & 0.27 & 4.45 & 99.99 & 1.5 & 0.47 &
3.67 & 99.99\tabularnewline
                       & 1.6 & 0.81 & 2.92 & 99.99 & 1.6 & 1.19 & 2.48 & 99.99\tabularnewline
                       & 2.0 & 4.65 & 1.39 & 99.96 & 1.9 & 4.55 & 1.40 & 99.96\tabularnewline
                       & 2.5 & 10.85 & 0.98 & 99.90 & 2.3 & 10.46 & 0.99 & 99.92\tabularnewline
$(8,0)$   & 1.3 & 0.06 & 5.84 & 99.99&1.3 & 0.12 & 5.39 &
99.99\tabularnewline
          & 1.6 & 2.13 & 1.94 & 99.99 & 1.6 & 2.51 & 1.81 & 99.99\tabularnewline
          & 2.0 & 7.01 & 1.18 & 99.99& 2.0 & 8.19 & 1.11 & 99.99\tabularnewline
          & 2.5 & 14.00 & 0.90 &99.95& 2.5 & 16.62 & 0.83 & 99.95\tabularnewline
$(1,0)$   & 1.0 & 2.32 & 1.90 & 99.11  & 1.0 & 0.73 & 3.09 &
99.35\tabularnewline
          & 1.1 & 20.33 & 0.84 & 98.15 & 1.1 & 16.02& 0.92 & 98.09\tabularnewline
          & 1.2 & 56.70 & 0.60 & 97.31 & 1.2 & 52.46 & 0.61 & 97.23\tabularnewline
          & 1.3 & 109.41 & 0.48 & 96.49 & 1.3 & 108.67 & 0.48 & 96.47
 \tabularnewline
\bottomrule[1pt]\bottomrule[1pt]
\end{tabular}
  \end{table}

\section{DISCUSSIONS AND CONCLUSIONS}\label{Seccnclsn}

In this work, we have performed a systematic investigation of the
possible deuteron-like states composed of a pair of doubly charmed
spin-$\frac{1}{2}$ baryons or one doubly charmed baryon and one
doubly charmed antibaryon. In the spin-triplet sector we take into
account mixing between the ${}^3S_1$ and ${}^3D_1$ channels. The
present formalism can also be extended to the loosely bound systems
composed of one spin-$\frac{1}{2}$ and one spin-$\frac{3}{2}$ or two
spin-$\frac{3}{2}$ baryons.

For the spin-triplet $B_{cc}B_{cc}$ systems, we obtain two loosely
bound states for $(R,I) = (\bar{3},1/2)$ and $(\bar{3},0)$. Their
binding energies are from a few MeV to tens of MeV and
root-mean-square radii from $1$ fm to a few fm for the cutoff around
$1.2 - 1.5$ GeV. They are good candidates of the molecular states.
In the spin-singlet sector, the potentials are not strong enough to
form bound states for $(R,I) = (6,1)$, $(6,1/2)$ and $(6,0)$ with a
reasonable cutoff value.

For the $B_{cc}\bar{B}_{cc}$ systems, the spin-singlet and
spin-triplet cases are similar. Very interestingly, we obtain
loosely bound states for the spin-singlet and spin-triplet systems
with $(R,I) = (8,1)$, $(8,1/2)$ and $(8,0)$. They have binding
energies around a few MeV and root-mean-square radii around a few
fm. They are also very good candidates of the molecular states in
the framework of the one-boson-exchange-potential model. We also
notice that the contact interaction plays a minor role in the
formation of the bound states for the $B_{cc}\bar{B}_{cc}$ systems.
The $D$-wave probability is tiny for most of the spin-triplet
channels.

Theoretical explorations of the exotic states containing multiple
heavy quarks first appeared nearly three decades ago \cite{chao}.
Recently these charming states are gaining more and more interest.
In the past several years, many events with four heavy quarks
($QQ\bar{Q}\bar{Q}$) have been reported experimentally
\cite{Aaij:2011yc,Khachatryan:2014iia,Abazov:2015fbl,Kamuran,Maksat}.
There are heated theoretical discussions of the exotic resonances
containing four heavy quarks recently
\cite{Brambilla:2015rqa,Chen:2016jxd,Wu:2016vtq,Bai:2016int,Karliner:2016zzc,Wang:2017jtz}.
The $B_{cc}\bar{B}_{cc}$ molecular states may be produced at LHC in
the near future. Once produced, they may decay into very
characteristic final states containing one or two charmonia,
including (1) two charmonia plus one or more light mesons/photons;
(2) one charmonium and a $D^{(*)}\bar{D}^{(*)}$ pair; (3) one
charmonium plus some photons or light mesons etc. They may also
decay into many light mesons or several hard photons. The
$B_{cc}\bar{B}_{cc}$ molecular states lie close to the mass
threshold of two doubly charmed baryons, which provides a clue to
identify them unambiguously. For example, these molecular states may
appear around $7\sim 7.5$ GeV depending on the mass of $\Xi_{bc}$.
Similarly, we also expect $B_{bc}\bar{B}_{bc}$ and
$B_{bb}\bar{B}_{bb}$ types of molecular states. They may lie roughly
around 14 GeV and 20 GeV respectively, if we take the mass values of
$\Xi_{bc, bb}$ in Ref.
\cite{Brodsky:2011zs,Karliner:2014gca,Sun:2014aya,Sun:2016wzh}.

Although very difficult to generate experimentally,
the bound states of $\Xi_{cc}\Xi_{cc}$ might be stable once produced
because $\Xi_{cc}$ decays via weak interaction most likely. There
might exist a strong decay mode: $\XX \rightarrow
\Omega^{++}_{ccc}A_c$, where $A_c$ is a charmed baryon and
$\Omega^{++}_{ccc}$ is the triply charmed baryon. The mass
estimation of triply charmed baryon can be found in Ref.
\cite{Brown:2014ena,Martynenko:2007je}. Whether the above decay
mode exists or not depends on the masses of the $\Xi_{cc}$ and
$\Omega^{++}_{ccc}$.

\section*{ACKNOWLEDGMENTS}

L. Meng is very grateful to G.J. Wang, H.S Li and B. Zhou for very
helpful discussions. The authors thank Ulf-G. Mei{\ss}ner and J.-M. Richard for helpful comments. This project is supported by the National Natural Science Foundation of China under Grants NO. 11621131001, 11575008 and 973 program. This
work is also supported in part by the DFG and the NSFC through funds
provided to the Sino-German CRC 110 ``Symmetries and the Emergence
of Structure in QCD".


\begin{appendix}

\section{Definitions of some functions and Fourier transform formulae} \label{app_func}

The definitions of the functions $H_i$ are \cite{Li:2012bt},
\begin{align}
H_0(\L,m,r)&=Y(u r)-\f{\l}{u}Y(\l r)-\f{r\beta^2}{2u}Y(\l r),  &H_1(\L,m,r)&=Y(u r)-\f{\l}{u}Y(\l r)-\f{r\l^2\beta^2}{2u^3}Y(\l r), \n\\
H_2(\L,m,r)&=Z_1(u r)-\f{\l^3}{u^3}Z_1(\l r)-\f{\l
\beta^2}{2u^3}Y(\l r), &H_3(\L,m,r)&=Z(u r)-\f{\l^3}{u^3}Z(\l
r)-\f{\l \beta^2}{2u^3}Z_2(\l r),
\end{align}
where,
\begin{eqnarray*}
 \beta^2=\L^2-m^2,\quad u^2=m^2-Q_0^2,\quad\l^2=\L^2-Q_0^2,
\end{eqnarray*}
and
\begin{eqnarray*}
 Y(x)=\f{e^{-x}}{x},\quad Z(x)=\left(1+\f{3}{x}+\f{3}{x^2}\right)Y(x),\quad
 Z_1(x)=\left(\f{1}{x}+\f{1}{x^2}\right)Y(x),\quad Z_2(x)=(1+x)Y(x).
\end{eqnarray*}

In our case all heavy hadrons have the same masses, we have
\begin{eqnarray}
 Q_0^2 =\left(\sqrt{m_f^2+\bm{p}_f^2}-\sqrt{m_i^2+\bm{p}_i^2}\right)^2 \approx  {\left( \bm{p}_i+\bm{p}_f\right)^2\bm{Q}^2 \over {4m_{\X}^2}}.
 \end{eqnarray}
Thus $Q_0^2$ is a high-order term and can be directly dropped out.

Without the form factor, one makes Fourier transformation and
obtains
\begin{eqnarray}
&&\frac{1}{u^{2}+\bm{Q}^{2}}\rightarrow\frac{e^{-ur}}{4\pi r}=\frac{u}{4\pi}Y(ur), \\
&&\frac{\bm{Q}}{u^{2}+\bm{Q}^{2}}\rightarrow-i\nabla\left(\frac{u}{4\pi}Y(ur)\right)=i\frac{u^{3}}{4\pi}Z_{1}(ur)\bm{r},  \\
&&\frac{\bm{Q}^{2}}{u^{2}+\bm{Q}^{2}}\rightarrow-\frac{u^{3}}{4\pi}Y(ur)+\delta^{(3)}(\bm{r}), \label{ps} \\
&&\frac{Q_{i}Q_{j}}{u^{2}+\bm{Q}^{2}}\rightarrow-\frac{u^{3}}{12\pi}\left[Z(ur)k_{ij}+Y(ur)\delta_{ij}\right]+\frac{1}{3}\delta^{(3)}(\bm{r})\delta_{ij},\label{vector}
\end{eqnarray}
where $k_{ij}=3\f{r_ir_j}{r^2}-\delta_{ij}$. Clearly, there exist
terms with a delta function $\delta^{(3)}(\bm{r})$ in
Eqs.~(\ref{ps}-\ref{vector}). In the current work, we call these
terms the contact interaction or delta interaction.
The very short-range interactions accounted by
the heavier-meson exchange are not taken into
account in the current analysis. In Ref.\cite{Dai:2017ont}, the short-range
annihilation force is introduced by fitting the data for the
nucleon-antinucleon system. However, introducing such short-range
interaction is not feasible for the $\Xi_{cc}\bar{\Xi}_{cc}$ systems
due to the lack of the experimental data. Luckily, the
$\Xi_{cc}\bar{\Xi}_{cc}$ annihilation force is of extremely short
range around $0.02$ fm. We are mainly interested in the loosely
bound molecular states which should not depend sensitively on the
short-range dynamics.

After introducing the form factor, the Fourier transformation
formulae read
\begin{eqnarray}
&&\f{1}{u^2+\bm{Q}^2}\mathcal{F}^2(Q)\rightarrow\f{u}{4\pi}H_0(\L,m,r),\nonumber \\
&&\f{\bm{Q}^2}{u^2+\bm{Q}^2}\mathcal{F}^2(Q)\rightarrow-\f{u^3}{4\pi}H_1(\L,m,r), \nonumber \\
&&\f{\bm{Q}}{u^2+\bm{Q}^2}\mathcal{F}^2(Q)\rightarrow\f{iu^3}{4\pi}\bm{r}H_2(\L,m,r),\nonumber \\
&&\f{Q_iQ_j}{u^2+\bm{Q}^2}\mathcal{F}^2(Q)\rightarrow-\f{u^3}{12\pi}\left[H_3(\L,m,r)
k_{ij}+H_1(\L,m,r)\delta_{ij}\right]. \label{FTformula}
\end{eqnarray}
One can also get the results without the contact interaction term by
a simple replacement in the above equations,
\begin{equation}
H_1(\Lambda, m, r) \rightarrow H_0(\Lambda, m, r).
\end{equation}

\begin{figure}[htp]
\centering
\begin{tabular}{ccc}
\includegraphics[width=0.3\textwidth]{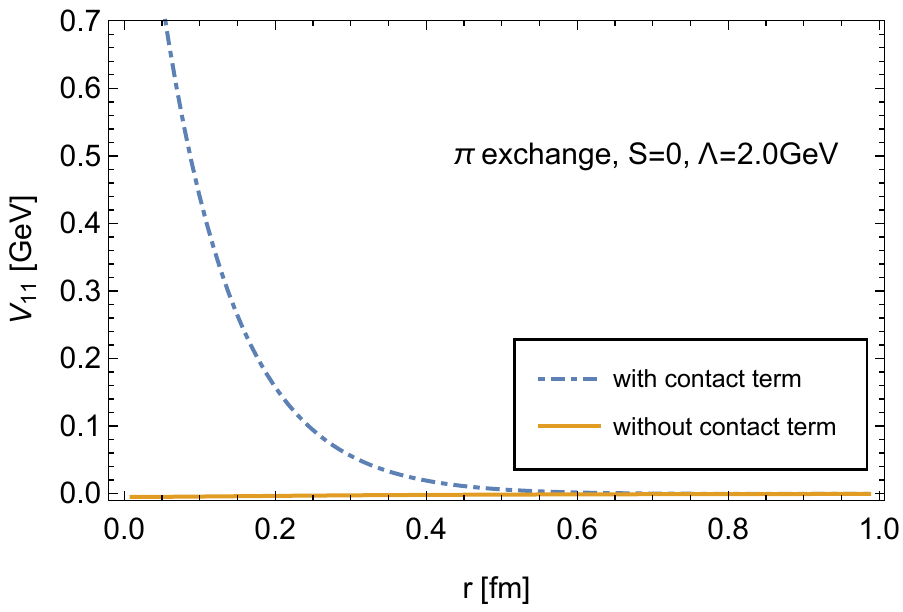}~&~
\includegraphics[width=0.3\textwidth]{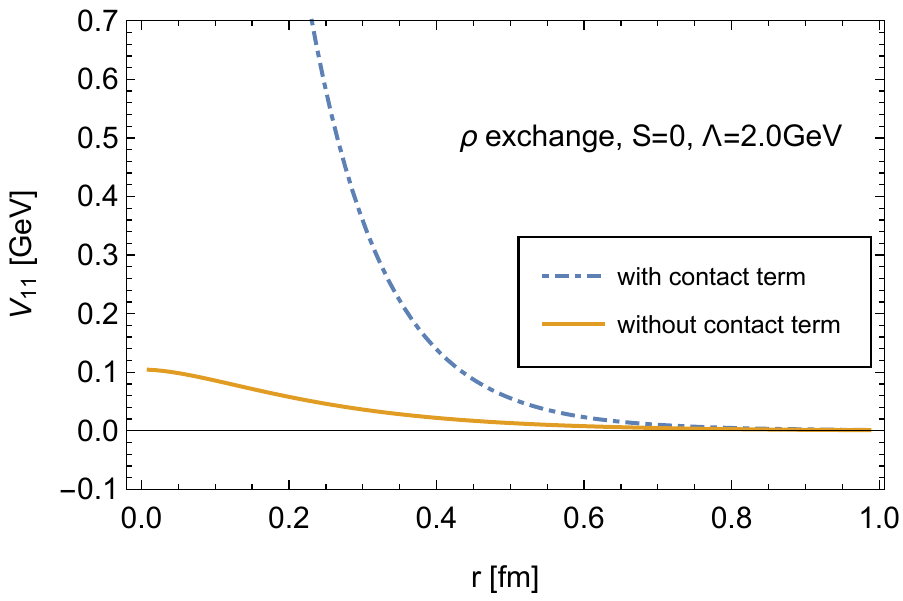}~&~
\includegraphics[width=0.3\textwidth]{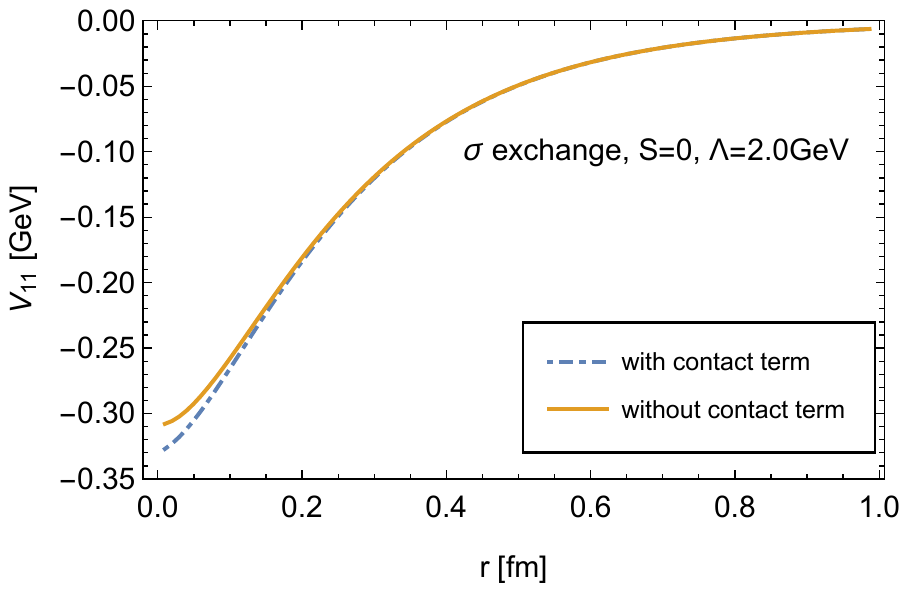}\\
\end{tabular}
\caption{(Color online). The potentials with/without the contact
terms for the ${}^1S_0$ channels. The isospin factors are set to 1.
}\label{pttldeltas0}
\end{figure}

\begin{figure}[htp]
\centering
\begin{tabular}{ccc}
\includegraphics[width=0.3\textwidth]{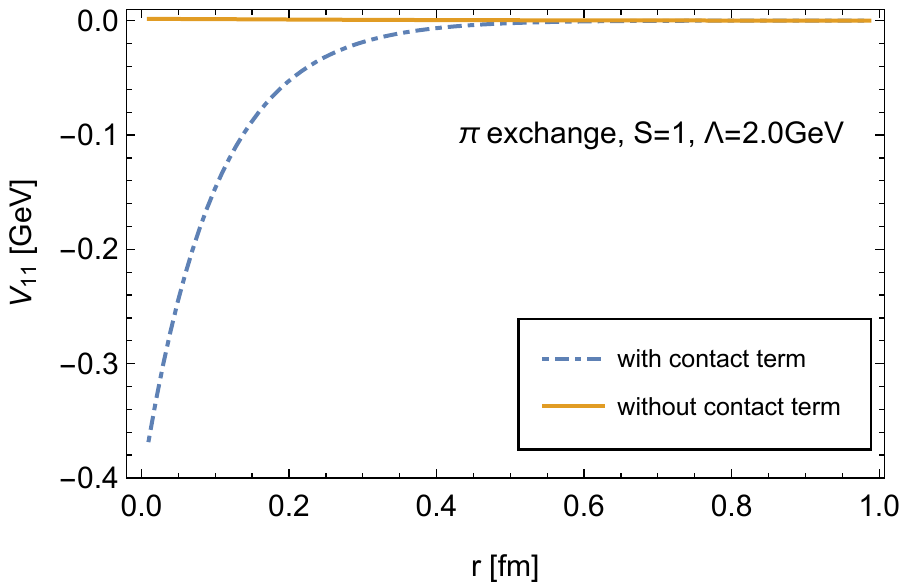}~&~
\includegraphics[width=0.3\textwidth]{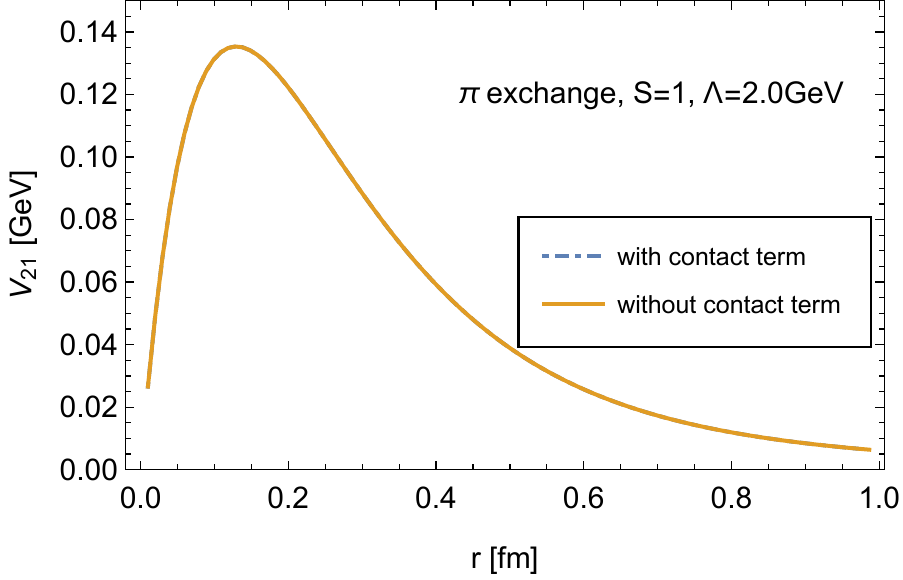}~&~
\includegraphics[width=0.3\textwidth]{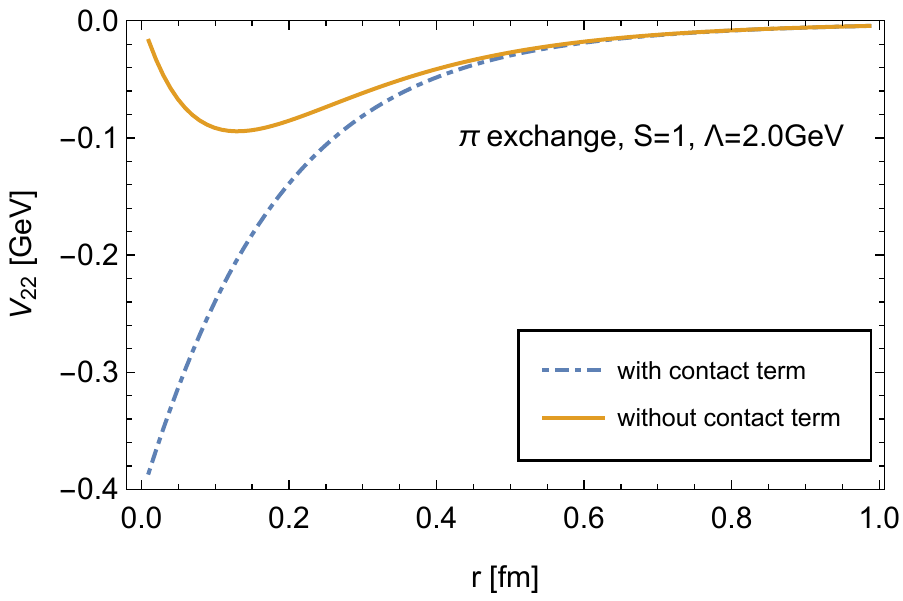}\\
\includegraphics[width=0.3\textwidth]{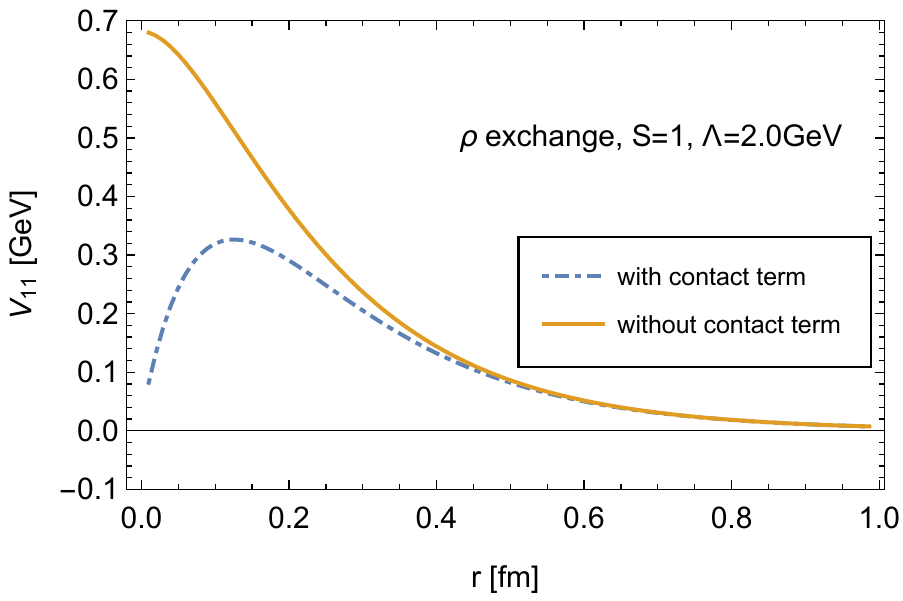}~&~
\includegraphics[width=0.3\textwidth]{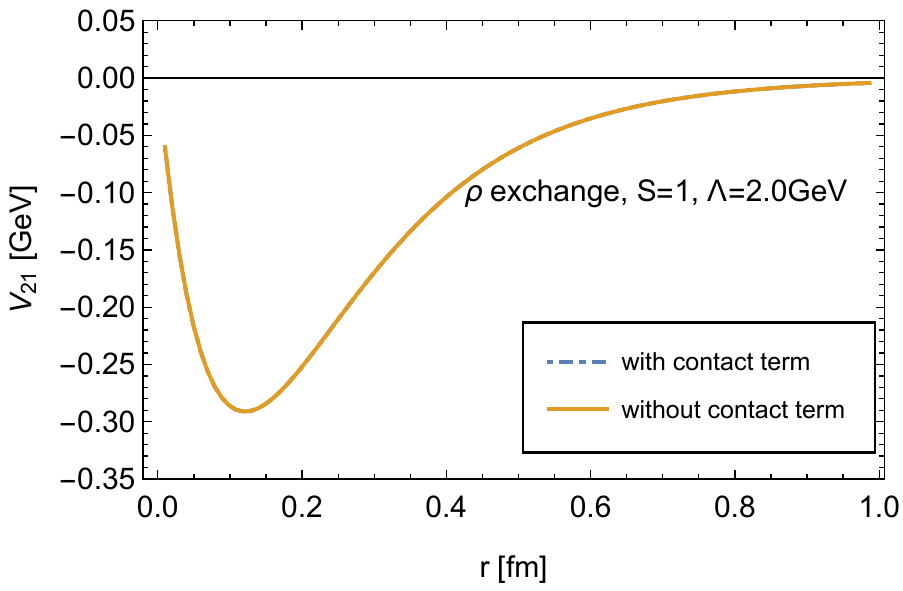}~&~
\includegraphics[width=0.3\textwidth]{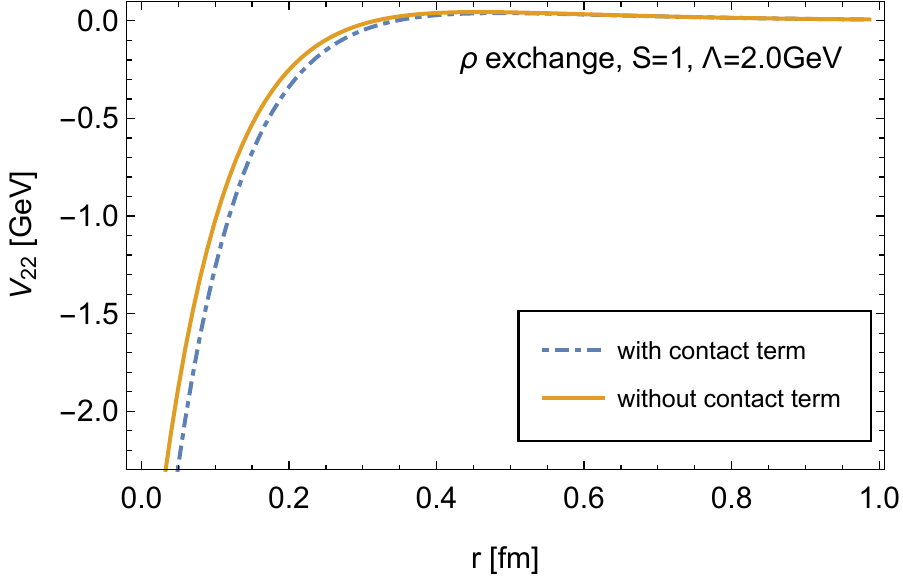}\\
\includegraphics[width=0.3\textwidth]{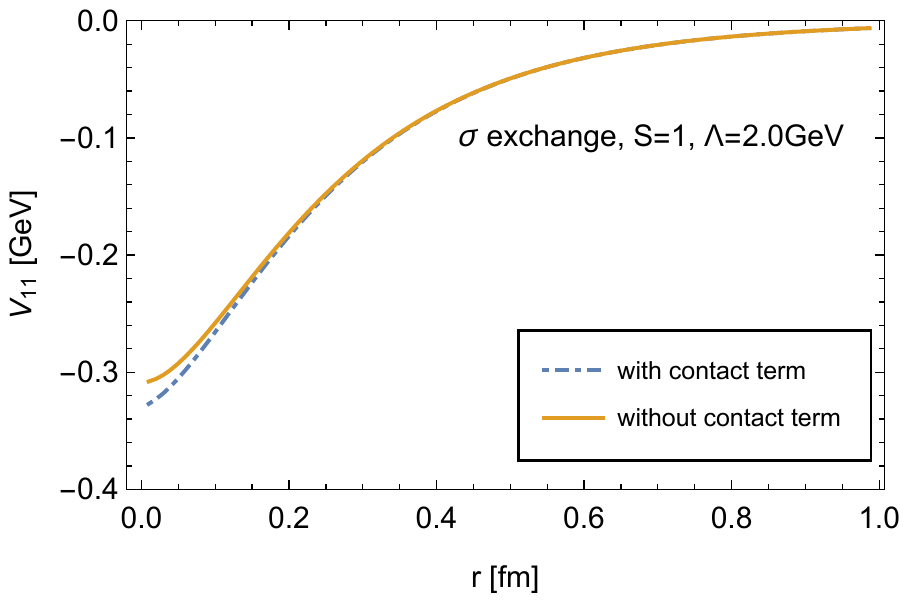}~&~
\includegraphics[width=0.3\textwidth]{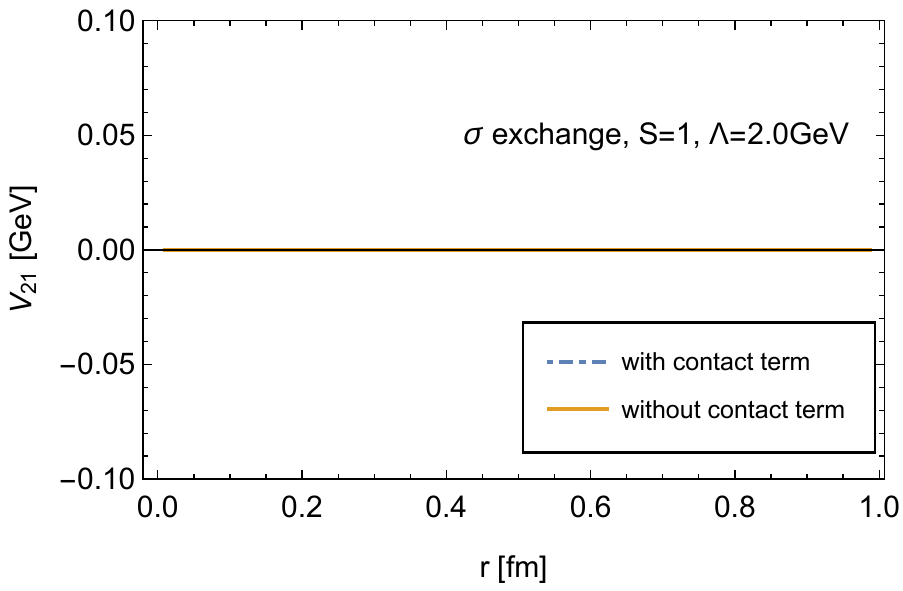}~&~
\includegraphics[width=0.3\textwidth]{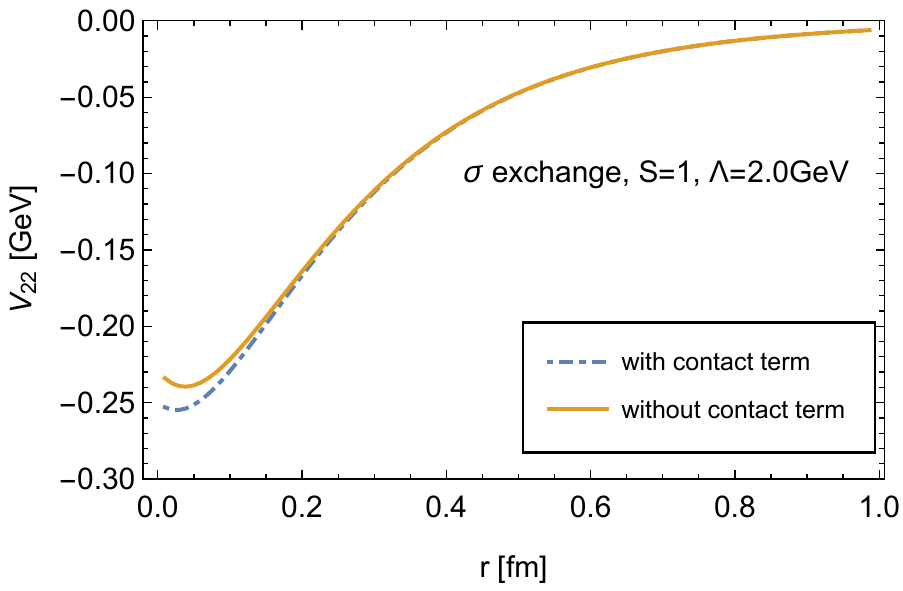}\\
\end{tabular}
\caption{(Color online). The potentials with/without the contact
terms for the coupled ${}^3S_1-{}^3D_1$ channels. The isospin
factors are set to 1. }\label{pttldeltas1}
\end{figure}

We show the interaction potentials both with and without the contact
interaction in Figs.~(\ref{pttldeltas0}-\ref{pttldeltas1}). We take
the $\pi$, $\rho$ and $\sigma$ exchange forces an example. The
isospin factors are set to $1$. From the plots, one can see clearly
that the contact interaction plays a minor role for the $\sigma$
exchange while its contribution is important in the range $r<0.4$ fm
for the $\pi$ and $\rho$ exchanges.

\section{Matrix elements of the operators}\label{app_matrix}

In the present work, we encounter the following operators,
\begin{itemize}
\item Spin-spin operator:
\begin{equation}
\bm{\s}_1\cdot\bm{\s}_2,
\end{equation}

\item Spin-orbit operator:
\begin{equation}
\bm{L}\cdot\bm{S},\quad\bm{L}\cdot\bm{S}_1,
\quad\bm{L}\cdot\bm{S}_2,
\end{equation}

\item Tensor operator:
\begin{equation}
S_{12}(\hat{r})=3(\bm{\sigma}_1 \cdot \hat{r})(\bm{\sigma}_2 \cdot
\hat{r})-\bm{\sigma}_1 \cdot \bm{\sigma}_2.
\end{equation}

\end{itemize}

For the spin-spin operator, one has
\begin{eqnarray}
\bm{\sigma}_1\cdot \bm{\sigma}_2 &=& 2\left( \bm{S}^2-\bm{S}_1^2-\bm{S}_2^2\right) \nonumber  \\
  &=& 2\left[S(S+1) - {3\over 2}\right].
\end{eqnarray}
The results are independent with the orbit angular momentum. For
spin-singlet and spin-triplet, the matrix elements of the spin-spin
interaction are -3 and 1 respectively.

For the spin-orbit operator one has,
\begin{eqnarray}
\bm{L}\cdot \bm{S}&=&{1\over 2}\left(\bm{J}^2-\bm{L}^2-\bm{S}^2\right)\\
&=& {1\over 2}\left[J_i(J_i+1)-L_i(L_i+1)-S_i(S_i+1)\right]
\end{eqnarray}
The results for $^1S_0$, $^3S_1$ and $^3D_1$ systems are 0, 0 and
-3/2 respectively. As for the $\bm{L}\cdot \bm{S}_{A(B)}$ type
interaction, the spin-orbit interaction vanishes for $^1S_0$,
$^3S_1$ systems. For the $^3D_1$ system, the spin wave function is
symmetric. The matrix elements of $\bm{L}\cdot \bm{S}_{A}$ and
$\bm{L}\cdot \bm{S}_{B}$ are the same, which are the half of the
matrix element of the operator $\bm{L}\cdot \bm{S}$.

The tensor operator is the scalar product of two rank-2 operator
$Y_{2,m}(\hat{r})$ and $T_{2,m}$,
\begin{equation}
S_{12}=\sum_{m=-2}^{2}4\sqrt{\frac{6\pi}{5}}T_{2,m}
Y^{*}_{2,m}(\hat{r}),
\end{equation}
where $Y_{2,m}(\hat{r})$ is the spherical harmonic function of
degree 2, and $T_{2,m}$ is rank-2 tensor operator constructed from
the total spin operator $\bm{S}$,
\begin{eqnarray}
&&T_{2,\pm2}=\frac{3}{8\pi}\left(S_{x}\pm iS_{y}\right)^{2}, \nonumber\\
&&T_{2,\pm1}=\mp\frac{3}{8\pi}\left[S_{z}\left(S_{x}\pm i S_{y}\right)+\left(S_{x}\pm i S_{y}\right)S_{z}\right],\nonumber \\
&&T_{2,0}=\sqrt{\frac{1}{6}}\frac{3}{4\pi}\left(3S_{z}^2-\bm{S}^{2}\right).
\end{eqnarray}
One can obtain the matrix elements of the tensor operator using the
Wigner-Echart theorem.

\end{appendix}

\vfil \thispagestyle{empty}

\end{document}